\documentclass[11pt,a4paper]{article}
\pdfoutput=1
\usepackage{jheppub}
\usepackage{amsmath,amssymb,bm}
\usepackage{subcaption}
\usepackage{hepunits}
\usepackage{braket}

\allowdisplaybreaks

\title{Higgs boson decay into four bottom quarks in the SM and beyond}

\author[a,b]{Jun Gao}
\emailAdd{jung49@sjtu.edu.cn}

\affiliation[a]{INPAC, Shanghai Key Laboratory for Particle Physics and Cosmology,
School of Physics and Astronomy, Shanghai Jiao Tong University, Shanghai 200240, China}
\affiliation[b]{Center for High Energy Physics, Peking University, Beijing 100871, China}

\abstract{
We present predictions for the Higgs boson decay into four bottom quarks in
the standard model and via light exotic scalars
retaining full bottom-quark mass dependence.
In the SM the decay can be induced either by the Yukawa couplings of bottom
quarks and top quarks or the electroweak couplings.
We calculate the partial decay width and various differential distributions
up to next-to-leading order in QCD.
We find large QCD corrections for decay via Yukawa couplings, as large as
90\% for the partial decay width, and reduced scale variations.
The results of this paper are therefore helpful for the measurement of this
multi-jets final state at future Higgs factory of electron-positron colliders.
We also propose several observables that can differentiate the SM
decay channel and the exotic decay channel and compare their next-to-leading
order predictions.
}

\begin{document}

\maketitle

\newpage

\section{Introduction}\label{sec:introduction}

The successful operation of the LHC and the ATLAS and CMS experiments
have led to the discovery of the Higgs boson and completion of the
standard model (SM) of particle physics~\cite{Aad:2012tfa, Chatrchyan:2012xdj}.
Further refined study at the LHC has revealed one essential
of the Higgs boson, Yukawa couplings of top quarks~\cite{Sirunyan:2018hoz, Aaboud:2018urx}
and bottom quarks~\cite{Aaboud:2018zhk, Sirunyan:2018kst}.
Precision test on properties of the Higgs boson including all its couplings
with standard model particles becomes one primary task of particle physics
at the high energy frontier.
In the SM the Higgs boson decays dominantly to hadronic final states which
are hard to access at hadron colliders due to huge QCD backgrounds.
That includes decay channels of a pair of bottom quarks or charm quarks,
a pair of gluons via top-quark loops, and four quarks via electroweak gauge
bosons, adding to a total decay branching ratio of about 80\%~\cite{1307.1347}.
To measure the Higgs properties with higher accuracy and to probe rare decay
modes of the Higgs boson, there have been a few proposals to build a future
lepton collider that can serve as a Higgs factory. 
These include ILC~\cite{Behnke:2013xla}, CEPC~\cite{CEPCStudyGroup:2018ghi},
CLIC~\cite{Lebrun:2012hj} and FCC-$ee$~\cite{Gomez-Ceballos:2013zzn}.
The proposed LHeC program can also produce abundant Higgs boson and measure
its couplings precisely~\cite{1206.2913,Klein:2018rhq}.
Certainly with the future Higgs factory, the precious hadronic decay channles
of the Higgs boson can be studied in details thanks to the clean environment~\cite{An:2018dwb}. 

Precision experiments require equally precision theoretical predictions.
To further scrutinize the SM and to look for possible new physics beyond,
it is necessary to calculate higher-order corrections to the production and
decay of the Higgs boson.
In this respect, there have been enormous advances in recent years.
For example, the next-to-next-to-next-to-leading order (N$^3$LO) QCD
corrections to Higgs boson production via gluon fusion in the heavy
top-quark limit~\cite{Anastasiou:2015ema, Mistlberger:2018etf} and
to Higgs boson production via vector boson fusion within the
structure function approach~\cite{Dreyer:2016oyx}, the next-to-next-to-leading
order (NNLO) corrections to Higgs boson production in association with a
jet in the heavy top-quark limit~\cite{Boughezal:2013uia, Chen:2014gva, Boughezal:2015dra, Boughezal:2015aha},
and the next-to-leading order (NLO) corrections to Higgs boson pair production with full
top-quark mass dependence~\cite{Borowka:2016ypz} have been known for some time.
The two-loop mixed QCD and electroweak corrections have also been calculated
recently for the associated production of Higgs boson and a $Z$ boson at
electron-positron colliders~\cite{Gong:2016jys, Sun:2016bel, Chen:2018xau}.
For decays of the Higgs boson, the partial width for $H \to b\bar{b}$
is known up to the next-to-next-to-next-to-next-to-leading order (N$^4$LO), in
the limit where the mass of the bottom quark is neglected~\cite{Baikov:2005rw,Davies:2017xsp,Herzog:2017dtz}.
The partial width for $H \to gg$ has been calculated to the N$^3$LO~\cite{Baikov:2006ch}
and N$^4$LO~\cite{Herzog:2017dtz} 
in the heavy top-quark limit.
We refer the readers to~\cite{Denner:2011mq, Spira:2016ztx} for a complete
list of relevant calculations.
At a more exclusive level, the fully differential cross sections
for $H \to b\bar{b}$ have been calculated to NNLO in~\cite{Anastasiou:2011qx, DelDuca:2015zqa}
and N$^3$LO in~\cite{Mondini:2019gid} for massless bottom quarks, and to NNLO in~\cite{Bernreuther:2018ynm}
with massive bottom quarks.

Besides, multi-jets final state, for instance, Higgs boson decays into
three or four QCD partons can also be explored at future Higgs factory.
There have been strong motivations for experimental searches of those
exclusive hadronic decay channels to look for exotic decays of the Higgs
boson induced by light states beyond the SM~\cite{0909.1521,1312.4992,Liu:2016zki,Liu:2016ahc}.
Among them there is the decay of the Higgs boson to four bottom quarks
which will be the main focus of this paper.
There have already been searches carried out by ATLAS collaboration at the LHC~\cite{1806.07355}
setting an upper limit of about 50\% on the decay branching ratio to four
bottom quarks.
On the theory side, the Higgs boson decays into four massless quarks via
electroweak gauge bosons have been calculated to NLO in both QCD and electroweak
couplings and have been implemented in the MC program \textsc{Prophecy4f}~\cite{Bredenstein:2006rh,Bredenstein:2006ha}.
There have also been recent predictions for Higgs boson decays
into three-jets final state~\cite{Li:2018qiy,1901.02253,1903.07277}
for the hadronic event shapes.
That is particularly interested for the probe of light-quark Yukawa
interactions~\cite{Gao:2016jcm}.
Very recently there exist a NNLO calculation for Higgs boson decays into
a pair of bottom quarks plus an additional jet for
massless bottom quarks~\cite{Mondini:2019vub}.   

In this paper, we present predictions for the Higgs boson decays into
four bottom quarks in the standard model and via light exotic scalars
retaining full bottom-quark mass dependence.
In the SM the decays can be induced either by the Yukawa couplings of bottom
quarks and top quarks or the electroweak couplings.
We calculate the partial decay width and various differential distributions
up to next-to-leading order in QCD.
We discuss details of the calculation in Section~\ref{sec:sm} and present
numerical results for the SM case.
We then move to discussion on exotic decays induced by new scalars 
including for the QCD corrections and comparisons to the SM case
in Section~\ref{sec:exo}.
We conclude in Section~\ref{sec:con}.

\section{Decay in the SM}\label{sec:sm}

\subsection{Decay via Yukawa interactions}
In the full theory with $n_f=6$ active flavor of quarks, the relevant interactions
include the Yukawa couplings of quarks and the QCD interactions,
\begin{equation}
\mathcal{L}_{int}=\mathcal{L}_{QCD}-\sum_{i=n_f-1}^{n_f}\frac{H}{v}m_i\bar \psi_{i}\psi_{i},
\end{equation}
We neglect quark masses except for top quark and bottom quark.
We use onshell scheme for renormalization of the quark or gluon fields and masses
except for masses in Yukawa couplings, when calculating the QCD radiative corrections.
The renormalization of the QCD coupling is carried out in $\overline {\rm MS}$ scheme
with top quark decoupled~\cite{Collins:1978wz}, namely the number of
light and heavy flavor, $n_l=5$ and $n_h=1$ respectively.
The corresponding renormalization constants at one-loop are,
\begin{eqnarray}\label{eq:ren}
Z_g&=&1+\frac{\alpha_S^{(n_l)}(\mu)}{4\pi}S_{\epsilon}\left\{-\frac{\beta_0^{(n_l)}}{\epsilon}
+\frac{2T_R}{3\epsilon}\sum_{i=n_f-n_h+1}^{n_f}\left(\frac{\mu^2}{m_i^2}\right)^{\epsilon}
\right\}, \nonumber\\
Z_2^G&=&1-\frac{\alpha_S^{(n_l)}(\mu)}{4\pi}S_{\epsilon}
\frac{4T_R}{3\epsilon}\sum_{i=n_f-n_h+1}^{n_f}\left(\frac{\mu^2}{m_i^2}\right)^{\epsilon},\nonumber\\
Z_2^{\psi_i}&=&1-\frac{\alpha_S^{(n_l)}(\mu)}{4\pi}S_{\epsilon}C_F\left\{
\frac{3}{\epsilon}+4\right\}
\left(\frac{\mu^2}{m_i^2}\right)^{\epsilon}, \nonumber\\
\frac{\delta m_i}{m_i}&=&-\frac{\alpha_S^{(n_l)}(\mu)}{4\pi}S_{\epsilon}
C_F\left\{{\frac{3}{\epsilon}
+4}\right\}\left(\frac{\mu^2}{m_i^2}\right)^{\epsilon},
\end{eqnarray}
with $S_{\epsilon}=(4\pi)^{\epsilon}/\Gamma(1-\epsilon)$, and
$\beta_0^{(n_l)}=(11C_A-4T_Rn_l)/6$.
Furthermore, since the relevant hard scale is much larger than the bottom quark mass,
we use the $\overline{\rm MS}$ running mass of bottom quark in Yukawa coupling,
that can be related to the input pole mass~\cite{hep-ph/0004189}.

\begin{figure}[ht]
\centering
\includegraphics[width=0.9\textwidth]{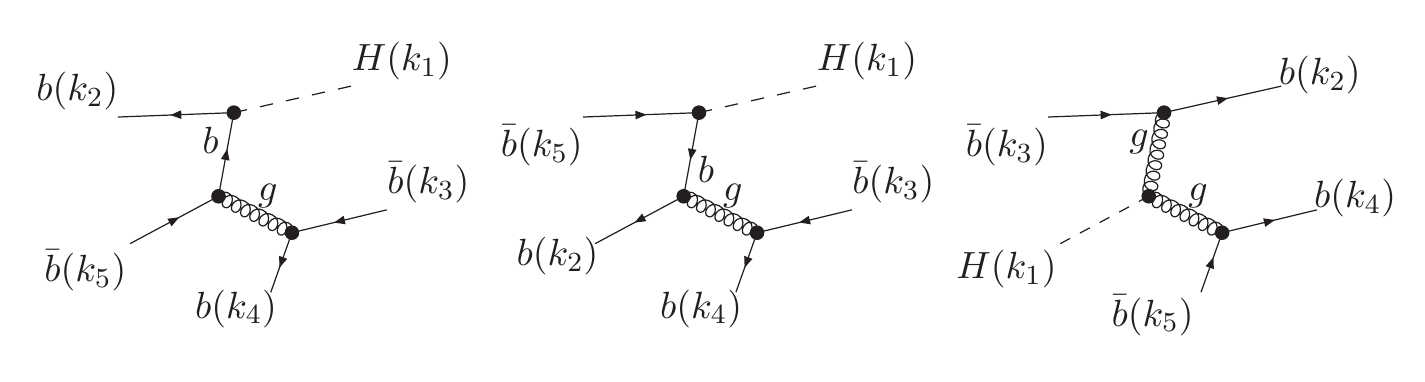}
\caption{
Feynman diagrams at leading order for the Higgs boson decaying into four bottom
quarks via Yukawa interactions.
Diagrams can be obtained with interchanges of identical particles are
not shown for simplicity. 
\label{fig:ff1}}
\end{figure}

Since the top quarks only appear as internal states, one can adopt an effective
theory by integrating out top quark, the interactions can be expressed as
\begin{equation}\label{eq:leff}
\mathcal{L}_{eff}=\mathcal{L}_{QCD}-\frac{H}{v}C_1G_{\mu\nu}^aG^{\mu\nu,a}
-\frac{H}{v}C_2\sum_{i=n_l}^{n_l}m_i\bar \psi_{i}\psi_{i},
\end{equation}
to the leading power of inverse of the top-quark mass.
The Wilson coefficients $C_1$ and $C_2$~\cite{Kataev:1981gr,Inami:1982xt,Dawson:1990zj,Djouadi:1991tka,Kataev:1993be,
hep-ph/9405325,Spira:1995rr,hep-ph/9708255} carry
further logarithmic dependence on top-quark mass and are given by
\begin{eqnarray}\label{eq:wil}
C_1&=&-\frac{\alpha_S^{(n_l)}(\mu)}{12\pi}\left\{1+
\frac{\alpha_S^{(n_l)}(\mu)}{4\pi}
11+\mathcal{O}(\alpha_S^2)
\right\}, \nonumber \\
C_2&=&1+\left(\frac{\alpha_S^{(n_l)}(\mu)}{4\pi}\right)^2\left\{{40\over 9}
-{16\over 3}\log{\mu^2\over m_t^2}\right\}+\mathcal{O}(\alpha_S^3).
\end{eqnarray}
The renormalization of QCD coupling and gluon field are done with pure $\overline {\rm MS}$
scheme with $n_l=5$ light flavors.
Again we use onshell scheme for renormalization of bottom quark field and mass, apart from using $\overline{\rm MS}$ running mass in the Yukawa coupling.
That is equivalent to set $n_h=0$ in Eq.~(\ref{eq:ren}). 
Moreover, in the effective theory, there will be operator mixing between the last
two terms in Eq.~(\ref{eq:leff}) requiring renormalization of the Wilson coefficients,
\begin{equation}
C_1^0=Z_{11}C_1, \quad \quad C_2^0=Z_{21}C_1+C_2,
\end{equation}
with the one-loop renormalization constants in $\overline{\rm MS}$ scheme given by~\cite{hep-ph/9708255},
\begin{eqnarray}
Z_{11}&=&1-\frac{\alpha_S^{(n_l)}(\mu)}{4\pi}S_{\epsilon}\frac{2\beta_0^{(n_l)}}{\epsilon},\nonumber\\
Z_{21}&=&-\frac{\alpha_S^{(n_l)}(\mu)}{4\pi}S_{\epsilon}\frac{12C_F}{\epsilon}.
\end{eqnarray}
Under the effective theory the leading order (LO) Feynman diagrams for the Higgs
boson decaying into four bottom quarks are shown in Fig.~\ref{fig:ff1}.
Those diagrams can be obtained with interchanges of identical particles
are not shown for simplicity.
In this calculation the squared amplitudes needed for a next-to-leading order
QCD calculation are generated automatically with program GoSam~2.0~\cite{1404.7096},
including for the one-loop virtual corrections and the real corrections.
Reduction of loop integrals are performed with Ninja~\cite{1203.0291,1403.1229}
and scalar integrals are calculated with OneLOop~\cite{0903.4665,1007.4716}.
We do not show here the one-loop virtual or real radiation diagrams which are
lengthy.
We use the dipole subtraction method with massive quarks~\cite{hep-ph/0201036} for
handling the QCD real corrections.

\subsection{Decay via electroweak interactions}
\begin{figure}[ht]
\centering
\includegraphics[width=0.9\textwidth]{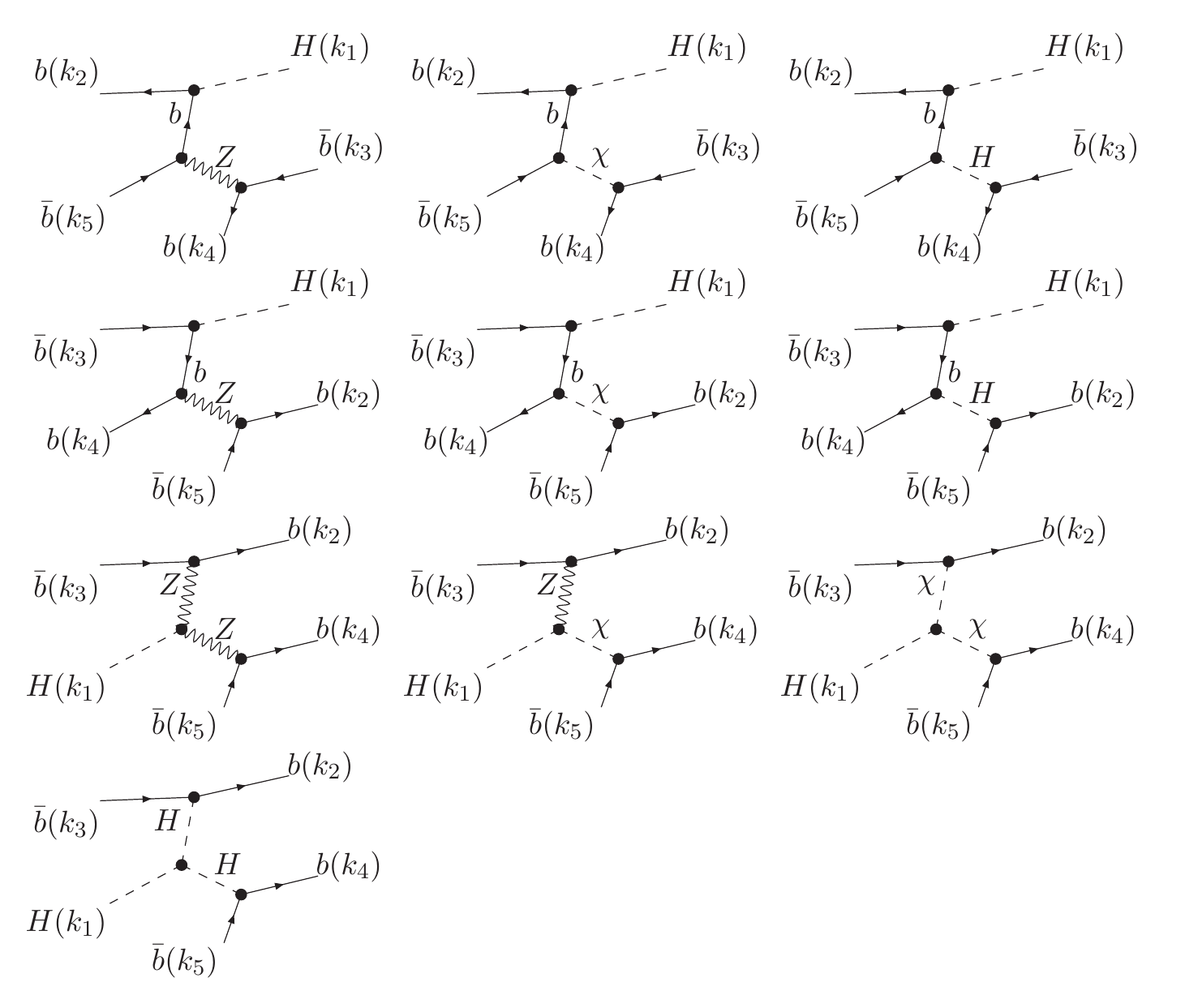}
\caption{
Similar to Fig.~\ref{fig:ff1} 
for Feynman diagrams of the Higgs boson decaying into four bottom
quarks via electroweak interactions.
\label{fig:ff2}}
\end{figure}

Within the standard model, the Higgs boson can also decay into four bottom quarks
through a cascade decay with an onshell $Z$ boson, $H\rightarrow Z(b\bar b) b\bar b$.
The corresponding branching ratio turns to be comparable to the one as induced by
the Yukawa couplings of bottom quarks and top quarks.
At leading order the relevant Feynman diagrams in Feynman-'t Hooft gauge
are shown in Fig.~\ref{fig:ff2}.
Those diagrams mediated by $Z$ boson and goldstone bosons $\chi$ must be considered
together to form a gauge invariant set.
Furthermore, we also include the diagrams mediated by Higgs bosons though
their contributions are small, since
we would like to keep full bottom quark mass dependence.
In principle there are also Feynman diagrams mediated by photons.
We do not consider them here since there at next-to-leading order in QCD
one will also need to include one-loop QED corrections of decay via Yukawa
couplings for consistency.

As mentioned earlier dominant contributions to decay via electroweak couplings arise
from the resonance region, namely one of the $b\bar b$ pairs lies at $Z$
boson mass pole.
Thus one must include finite width effects of the electroweak gauge bosons which may
violate gauge symmetry since that will mix contributions from various orders of the
EW couplings.  
In order to preserve gauge symmetry especially at one-loop level in QCD, we use the
complex mass scheme~\cite{hep-ph/0505042}.
There the masses of $W$ and $Z$ bosons and the electroweak couplings
are complex numbers depending on the width of $W$ and $Z$ bosons, and
the Lagrangian are manifestly gauge invariant.
The squared amplitudes needed for the next-to-leading order QCD calculation
are again generated automatically with GoSam~2.0~\cite{1404.7096} and are checked
against MadGraph5\_aMC@NLO~\cite{1405.0301}.
Good agreements are found between the two programs.

\subsection{Inclusive decay rate}

Similar to the total hadronic width, the inclusive decay width of the Higgs boson
to four bottom quarks in the limit of infinite top-quark mass includes contributions
from the bottom-quark Yukawa coupling, the gluon effective coupling, and their
interferences.
It can be expressed as~\cite{hep-ph/0503172}
\begin{eqnarray}
\Gamma_{4b,yuk}&=&\left(\frac{\alpha_S(\mu)}{2\pi}\right)^2
\Big\{A_{b\bar b}[\Delta_{b\bar b}(x)(1+\delta_{b\bar b}(x))C_2^2
+\Delta_{bg}(x)(1+\delta_{bg}(x))C_1C_2]\nonumber \\
&&+A_{gg}[\Delta_{gg}(x)(1+\delta_{gg}(x))C_1^2]\Big\},
\end{eqnarray}
with
\begin{equation}
A_{b\bar b}=\frac{3M_H}{8\pi v^2}\overline {m}_b^2(\mu),\qquad A_{gg}=\frac{4M_H^3}{2\pi v^2},
\qquad x=m_b^2/M_H^2,
\end{equation}
and $C_1(\mu)$, $C_2(\mu)$ as given in Eq.~(\ref{eq:wil}). 
The leading-order form factors $\Delta_{ij}$ carry up to quadratic dependence on
logarithm of the bottom quark mass.
Scale dependent terms in the next-to-leading order corrections $\delta_{ij}$ can
be obtained through renormalization scale invariance, e.g.,
\begin{eqnarray}
\delta_{b\bar b}(x)&=&\frac{\alpha_S(\mu)}{2\pi}\left[(2\beta_0+3C_F)\ln(4\mu^2/M_H^2)
+a_{b\bar b}(x)\right],\nonumber \\
\delta_{b g}(x)&=&\frac{\alpha_S(\mu)}{2\pi}\left[(3\beta_0+3C_F)\ln(4\mu^2/M_H^2)
+a_{b g}(x)\right],\nonumber \\
\delta_{g g}(x)&=&\frac{\alpha_S(\mu)}{2\pi}\left[(4\beta_0)\ln(4\mu^2/M_H^2)
+a_{g g}(x)\right].
\end{eqnarray}

Furthermore, the contributions via Higgs boson decaying into electroweak gauge bosons
can be written as
\begin{equation}
\Gamma_{4b,ew}=A_{ZZ}\Delta_{ZZ}(x)(1+\delta_{ZZ}(x)),
\end{equation}
with
\begin{equation}
A_{ZZ}=\frac{32M_Z^4M_H}{\pi^3v^4},\qquad \delta_{ZZ}=\frac{\alpha_S(\mu)}{2\pi}[a_{ZZ}(x)].
\end{equation}
In this case both $\Delta_{ZZ}(x)$ and $a_{ZZ}(x)$ depend on the bottom-quark mass weakly.  
We do not consider the interferences between decay induced by
Yukawa couplings and via the electroweak gauge bosons.

\begin{table}[h!]
\centering
\begin{tabular}{|cc|cccccc|cc|}
\hline
$m_b$ (GeV)  & $x\, (10^{-3})$ & $\Delta_{b\bar b}$ & $\Delta_{bg}$ & $\Delta_{gg}$ 
& $a_{b \bar b}$ & $a_{bg}$ & $a_{gg}$  & $\Delta_{ZZ}$& $a_{ZZ}$ \tabularnewline
\hline                      
 4.2  & 1.129 & 7.32 & -144.0 & 1.160 & 45.2 & 56.9 & 57.8 & 0.1222&  5.64\tabularnewline
\hline                      
 4.4  & 1.239 & 6.80 & -133.3 & 1.094 & 45.2 & 56.0 & 56.7 & 0.1205& 5.80\tabularnewline
\hline                      
 4.6  & 1.354 & 6.32 & -123.4 & 1.032 & 45.1 & 55.2 & 55.7 &0.1188 & 5.97\tabularnewline
\hline                      
 4.8  & 1.474 & 5.89 & -114.7 & 0.976 & 45.0 & 54.5 & 54.8 &0.1170 & 6.14\tabularnewline
\hline                      
 5.0  & 1.600 & 5.49 & -106.7 & 0.922 & 44.9 & 53.8 & 53.9 &0.1152& 6.32\tabularnewline
\hline
 5.2  & 1.730 & 5.13 & -99.4 & 0.873 & 44.9 & 53.2 & 53.2 &0.1133& 6.50\tabularnewline
\hline
\end{tabular}
\caption{
 Numerical results of the LO form factors and their NLO corrections for representative
 values of the bottom-quark pole mass and with $M_H=125\,{\rm GeV}$. 
\label{tab:inc}}
\end{table}

\begin{figure}[ht]
\centering
\includegraphics[width=0.47\textwidth]{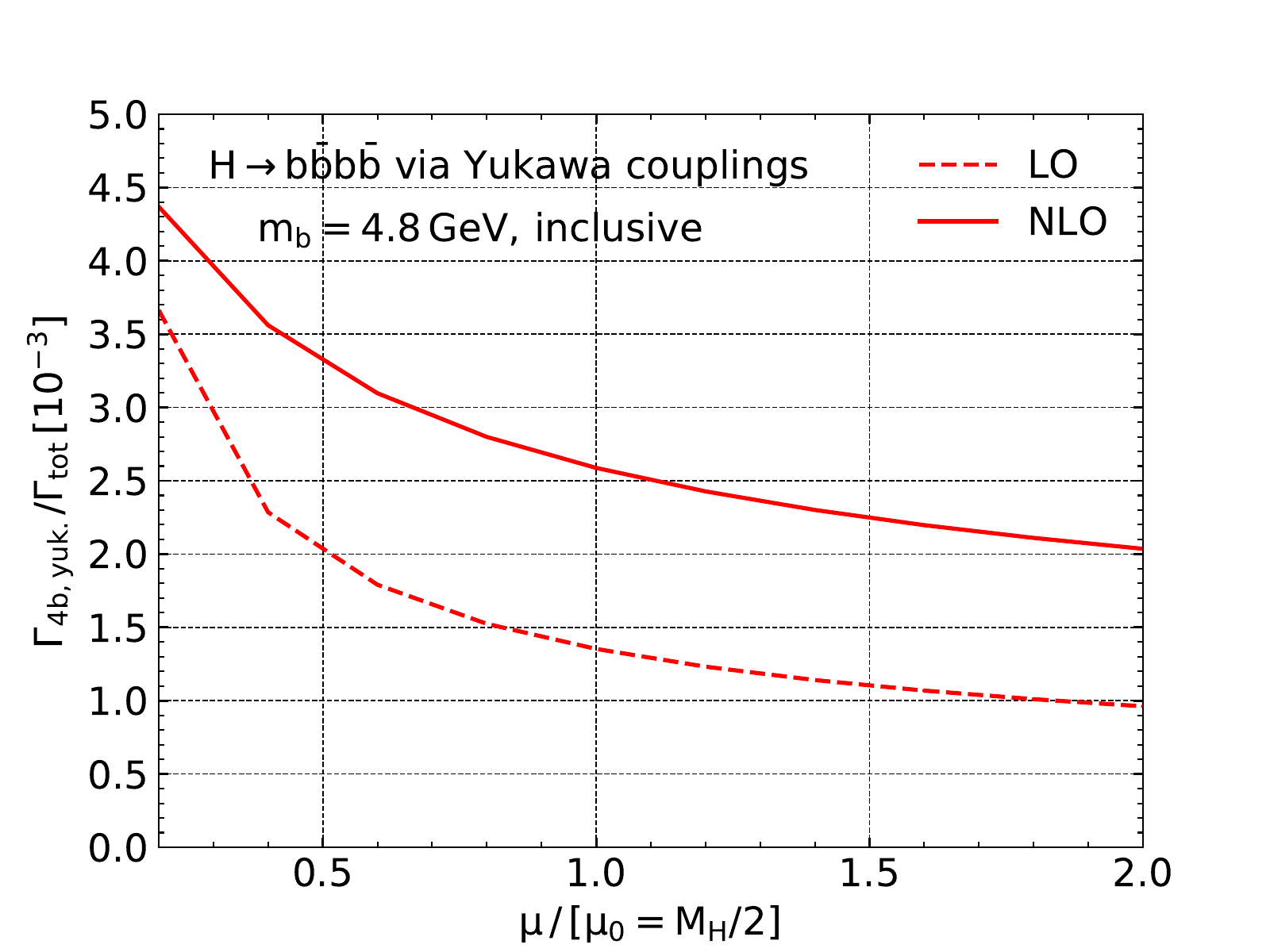}\hspace{0.1in}
\includegraphics[width=0.47\textwidth]{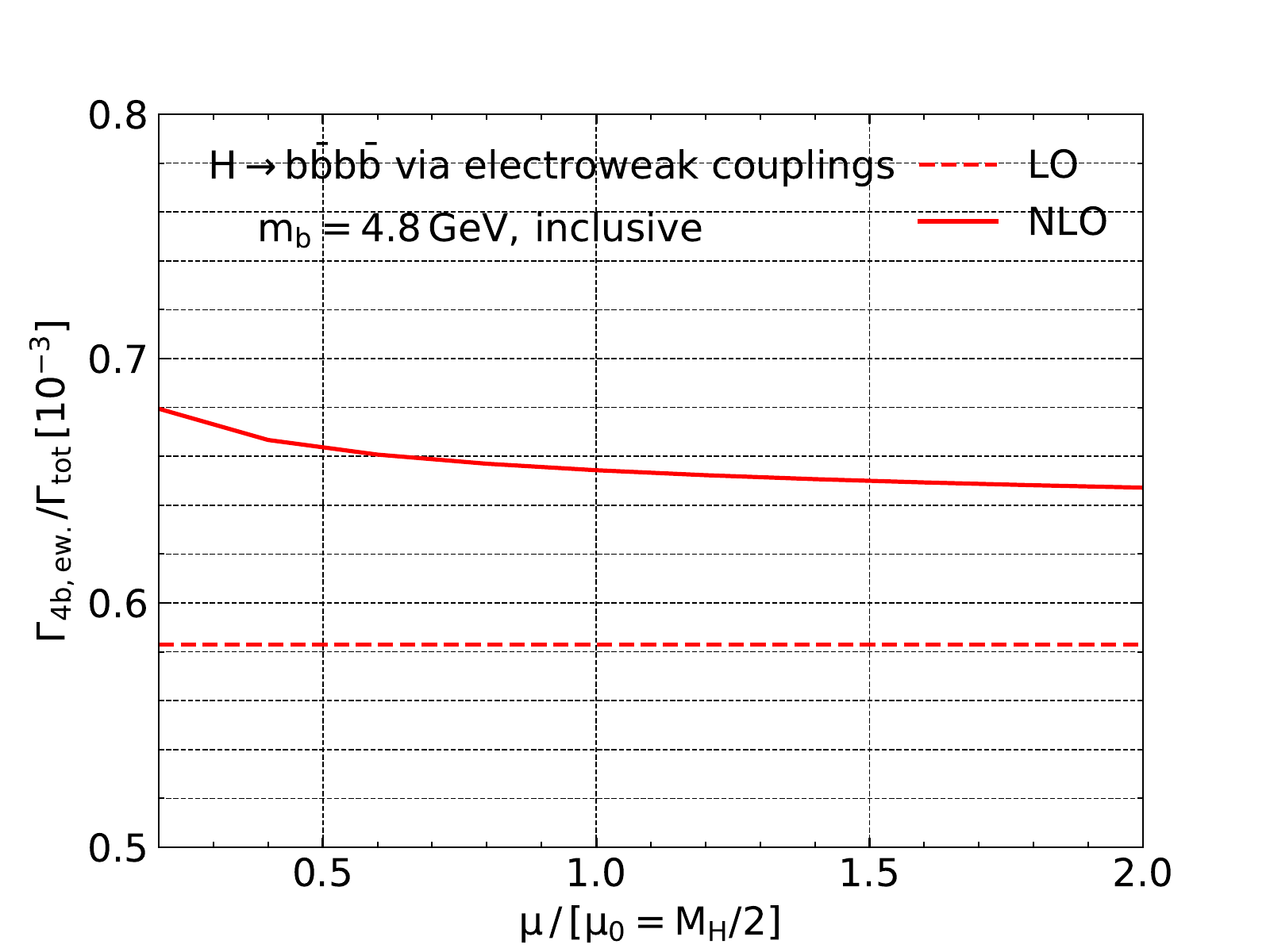}
\caption{
Decay branching ratio of the Higgs boson to four bottom quarks as a function
of the renormalization scale via either Yukawa couplings (left plot) or the electroweak
couplings (right plot), at both LO and NLO. 
\label{fig:inc}}
\end{figure}

Full mass dependence of factors $\Delta_{ij}$ and $a_{ij}$
can be complicated.
We provide their numerical values in Table.~\ref{tab:inc}
for several choices of the bottom-quark pole mass.
We set mass of the Higgs boson $M_H=125\,{\rm GeV}$, vacuum expectation value
$v=246.22\,{\rm GeV}$, and $\alpha_S(M_Z)=0.118$ in
all numerical calculations.
The from factors further depend on the masses of the electroweak gauge bosons, as
well as their width, which we set to~\cite{Tanabashi:2018oca}
\begin{align}
M_W=80.379\,{\rm GeV}, & \qquad M_Z=90.1876\,{\rm GeV},\nonumber \\
\Gamma_Z=2.4952\,{\rm GeV}, &\qquad \Gamma_W=2.085\,{\rm GeV}.
\end{align} 
Negative sign of $\Delta_{bg}$ shown in Table.~\ref{tab:inc} indicates
a constructive interference between the Feynman diagrams due to bottom-quark Yukawa
coupling and those induced by effective coupling with gluons.
We found large NLO QCD corrections for $\Delta_{b\bar b}$, $\Delta_{gg}$, and the
interference $\Delta_{bg}$ contributions, with mild dependence on the mass of the bottom quark.  
We further calculate the decay branching ratio of the four bottom quark channels
assuming a total width of the Higgs boson $\Gamma_{tot}$ of 4 MeV~\cite{1101.0593}.
We plot the decay branching ratio as a function of the renormalization scale in
Fig.~\ref{fig:inc} at both LO and NLO for a bottom-quark mass of 4.8 GeV.
The decay branching ratio due to Yukawa couplings can reach a few per mill
and receives large QCD corrections.
The LO prediction has a large scale uncertainty which is improved with the
NLO corrections.
The NLO prediction amounts to $2.59^{+0.7}_{-0.6}\times 10^{-3}$ if using a
central scale of $\mu_0=M_H/2$ and a conventional scale variation by a factor
of two.
The decay branching ratio via electroweak couplings of the bottom quarks
is smaller and receives mild QCD corrections.
The NLO prediction is $0.656^{+0.01}_{-0.01}\times 10^{-3}$ using the same
scale choice.
\begin{figure}[ht]
\centering
\includegraphics[width=0.6\textwidth]{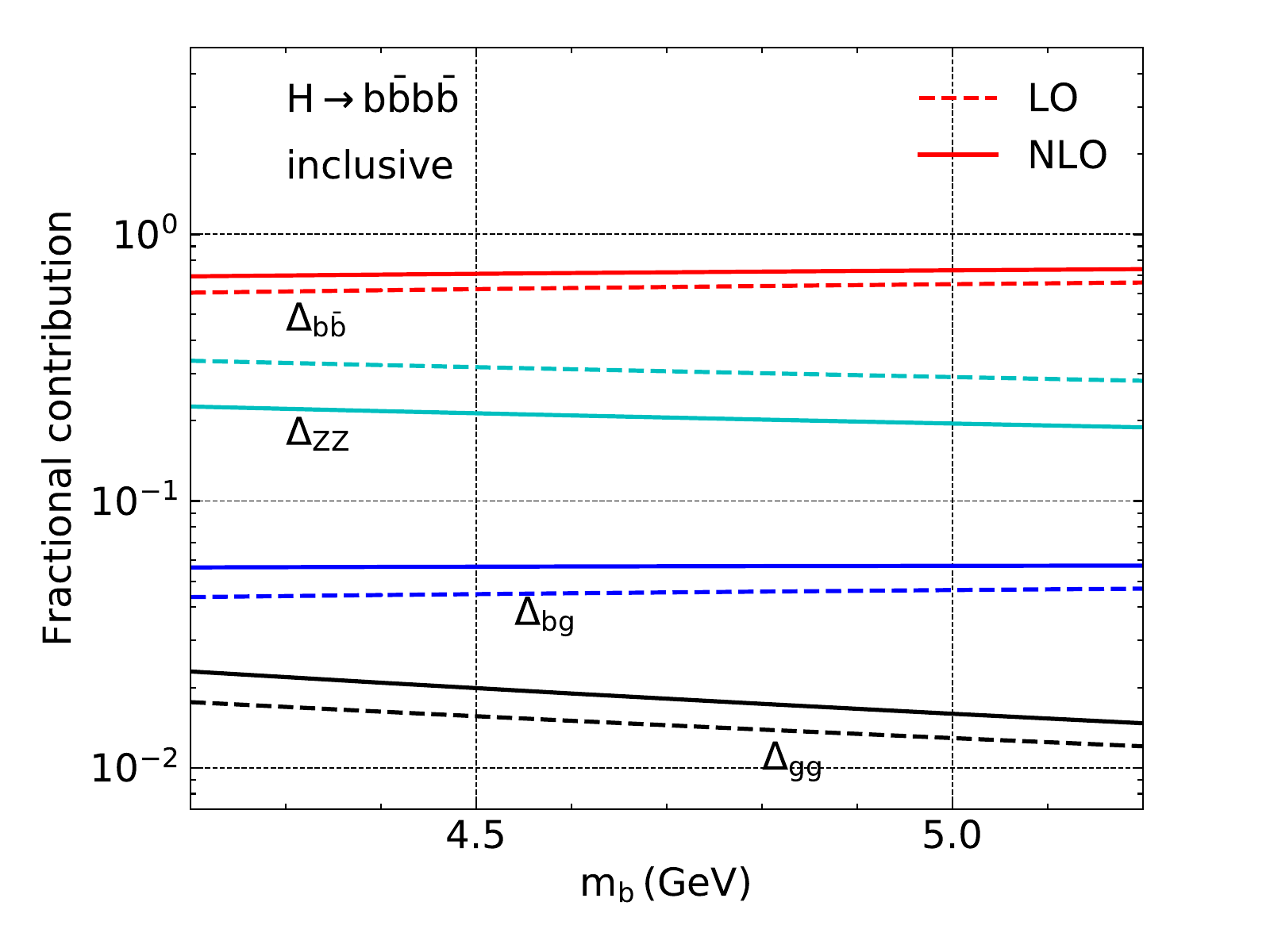}
\caption{
Fractional contributions to partial decay width of the Higgs boson
to four bottom quarks as a function of the
bottom quark mass, at both LO and NLO. 
\label{fig:frainc}}
\end{figure}

Fractional contributions from different terms to the total decay
branching ratio to four bottom quarks are shown
in Fig.~\ref{fig:frainc} as a function of the bottom quark mass.
Decay via electroweak couplings is sub-dominant but has larger contribution than
interference of the bottom-quark Yukawa coupling and the gluon effective coupling. 
Decay due to pure gluon effective coupling is at the level of a few percents
of the total decay branching ratio.

\subsection{Jet cross sections}

We consider final state with at least 4 $b$-tagged jets to separate from
other multi-parton hadronic decay modes, for instance, Higgs boson decaying
into a bottom quark pair plus two gluons or two light quarks. 
We use the $k_T$ jet algorithm~\cite{Catani:1991hj} with a resolution parameter $y$
varied between $10^{-3}$ and $0.5$.
The separation of any two clusters are calculated as
\begin{equation}
d_{ij}=\frac{2\,{\rm min}(E_i^2,E_j^2)}{Q^2}(1-\cos\theta_{ij}),
\end{equation}
with $Q^2=M_H^2/4$.
Constituents of jets are combined with $E$ scheme by directly summing
the four momentums.
Flavor of jets are defined by counting net $b$-quark numbers
within the jet, namely a jet with a $b$ quark and a $b$ anti-quark is considered
as light-flavor jet~\cite{0704.2999}.   

\begin{figure}[ht]
\centering
\includegraphics[width=0.47\textwidth]{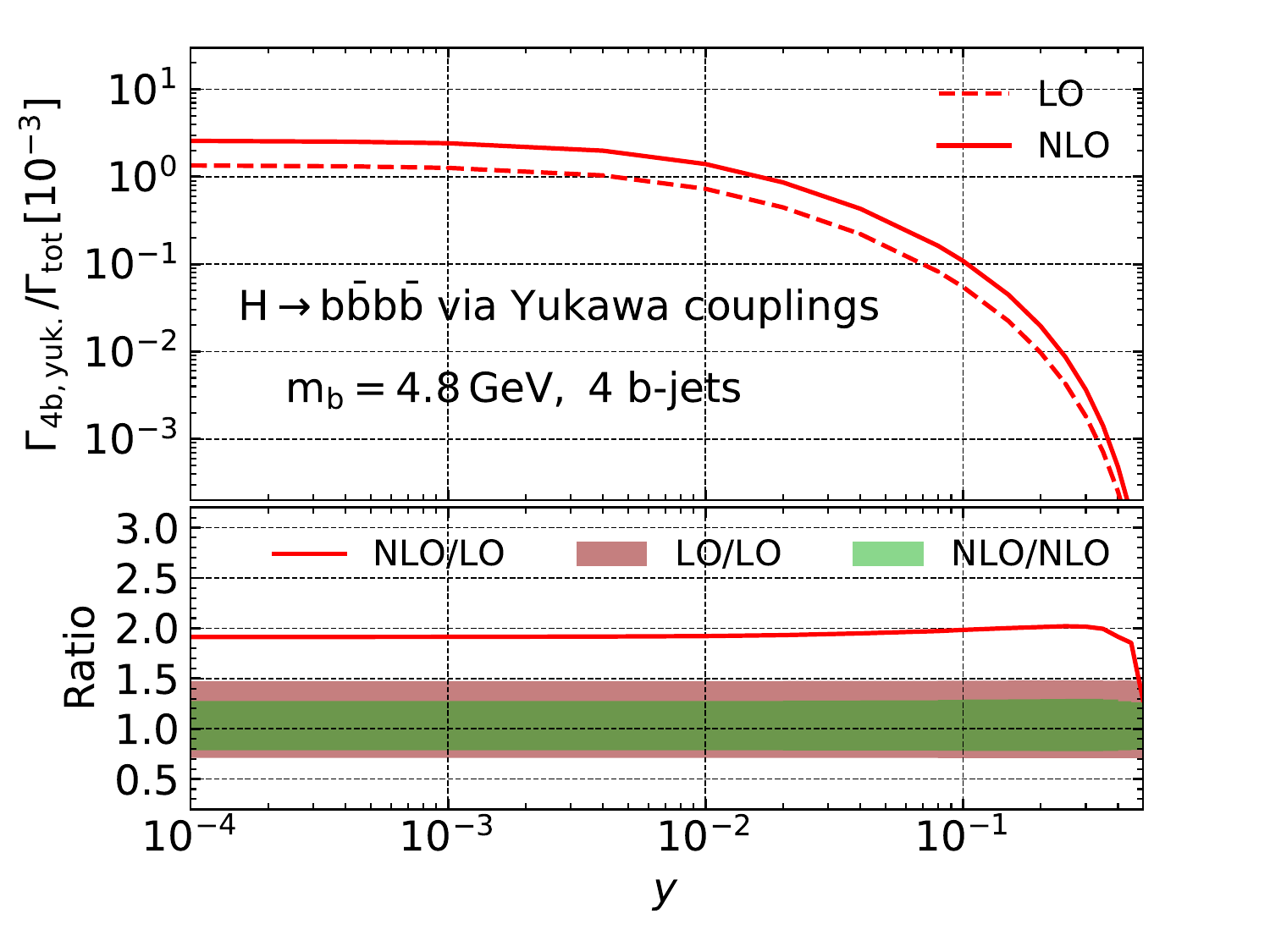}\hspace{0.1in}
\includegraphics[width=0.47\textwidth]{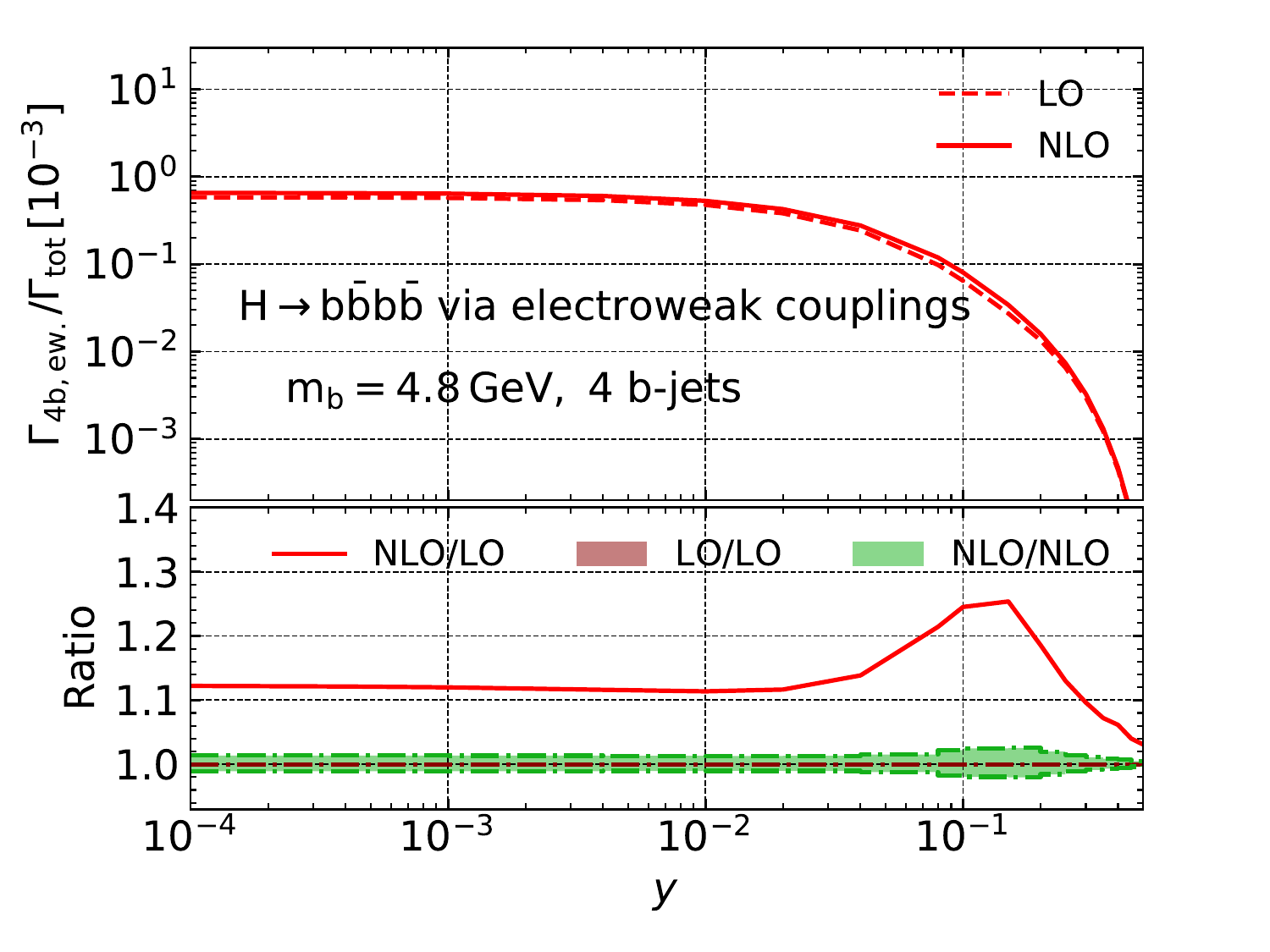}
\caption{
Decay branching ratio of the Higgs boson to four $b$-jets as a function
of the jet resolution parameter via either Yukawa couplings (left plot) or the electroweak
couplings (right plot), at both LO and NLO. 
\label{fig:jet}}
\end{figure}

\begin{figure}[ht]
\centering
\includegraphics[width=0.6\textwidth]{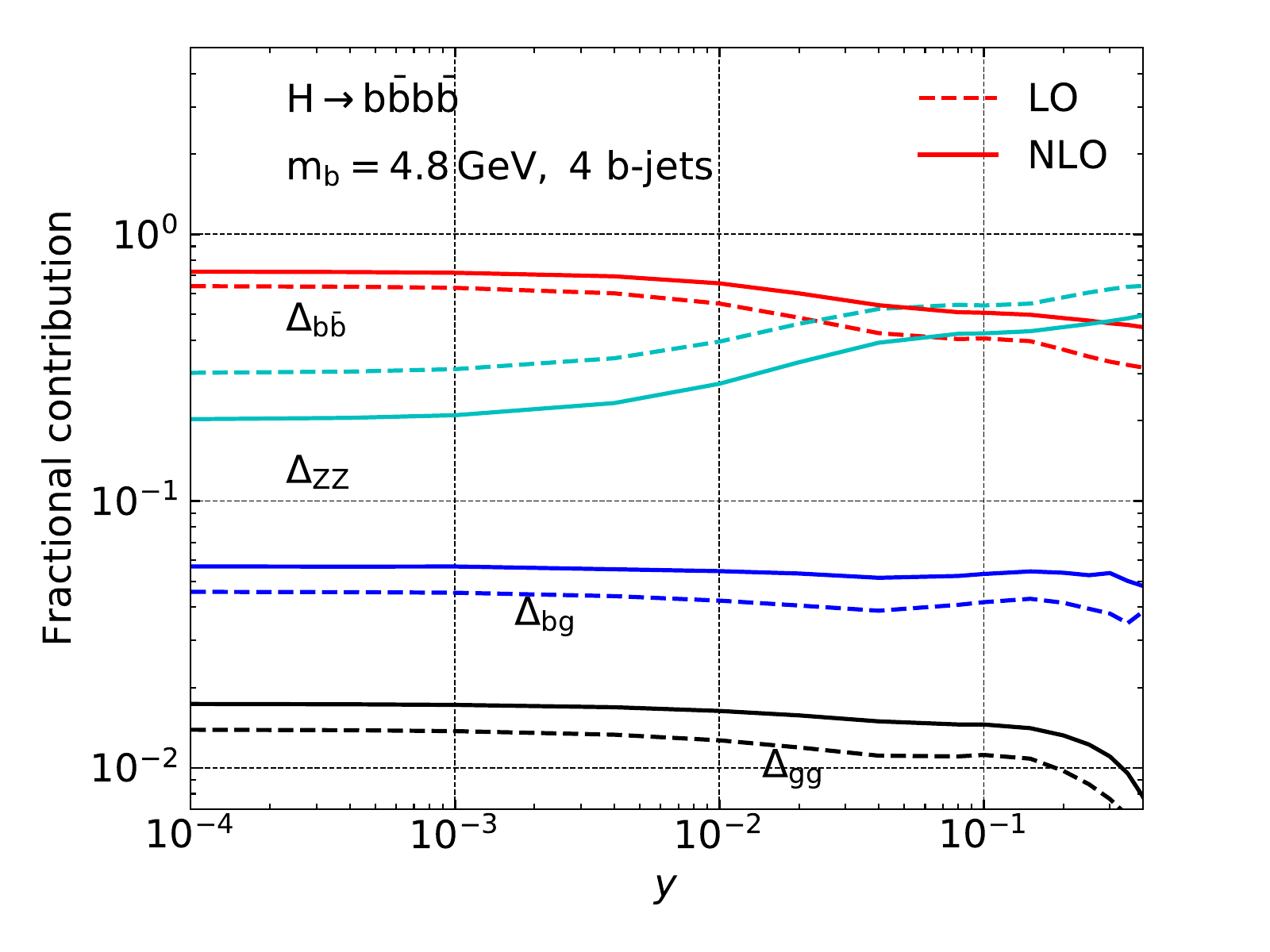}
\caption{
Fractional contributions to partial decay width of the Higgs boson
to four $b$-jets as a function of the jet resolution parameter,
at both LO and NLO. 
\label{fig:frajet}}
\end{figure}

We show the decay branching ratio to 4 $b$-jets as a function of
the jet resolution parameter in Fig.~\ref{fig:jet} via either Yukawa
couplings or electroweak couplings with a bottom-quark mass of
4.8 GeV.
In the upper panel we plot the branching ratio at LO and NLO.
In the lower inset we show several ratios including the NLO prediction to the LO one
for the nominal scale choice and the scale variations of the LO or NLO
predictions.
The jet rate approaches the inclusive rate of the decay when
$y$ goes to 0 since then the four bottom quarks are always fully resolved.
It decreases rapidly as the increasing of $y$ especially for the
case of decay via Yukawa couplings where the dominant contributions
are from quasi-collinear region of $b\bar b$ in the phase space.
The effects of QCD corrections are similar as in the inclusive
decay rate and exhibit mild dependence on the resolution parameter
except when close to the endpoint region of the phase space.
We found sizable NLO corrections and reduced scale variations
for decay via Yukawa couplings.
We further summarize the fractional contributions from different
channels to the jet rate in Fig.~\ref{fig:frajet}.
Contributions from $\Delta_{b\bar b}$ are always dominant while the
contributions from $\Delta_{ZZ}$ increase with the increasing of the resolution parameter.

\subsection{Event topology}

We consider several kinematic distributions of the final state $b$-jets,
that includes energy of individual jets, invariant mass and energy of $b$-jet pairs.
Jets are ordered according to their energies.
For invariant mass of $b$-jet pairs, we include the highest and lowest mass among all
combinations, $M_{bb,H}$, $M_{bb,L}$, the inclusive mass $M_{bb,inc}$ by counting
all possible combinations, and the mass asymmetry $\Delta M_{bb}$ that is the minimum of
absolute mass difference of two jet pairs for all possible divisions.
We define similar variables for energy of jet pairs, highest and lowest pair energy
$E_{bb,H}$ and $E_{bb,L}$, inclusive pair energy $E_{bb,inc}$, and the pair energy
asymmetry $\Delta E_{bb}$.
In case that the 4 $b$-jets arise from a cascade decay of two scalars with
identical masses, the mass or energy asymmetry peaks at zero values.       

\begin{figure}[ht]
\centering
\includegraphics[width=0.47\textwidth]{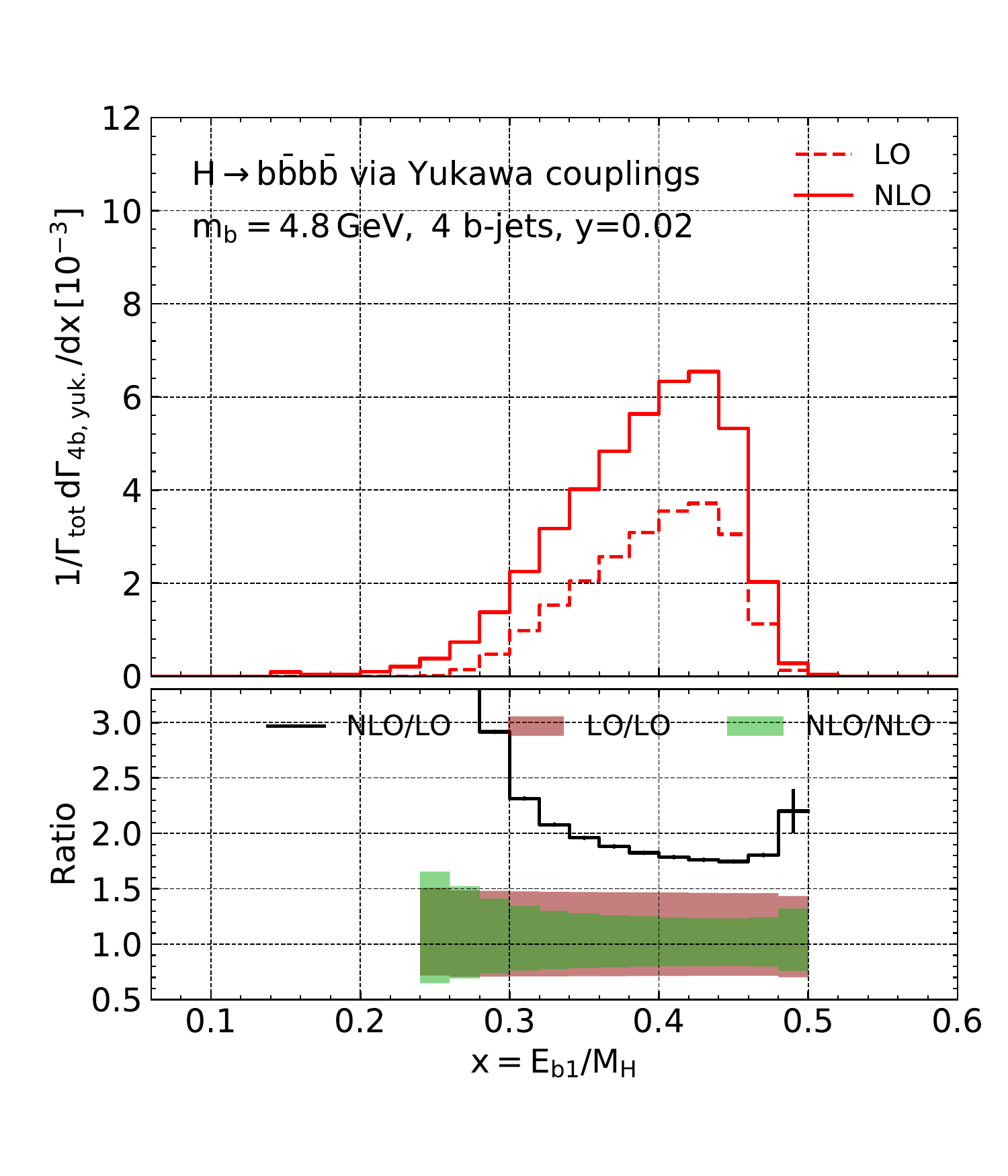}\hspace{0.1in}
\includegraphics[width=0.47\textwidth]{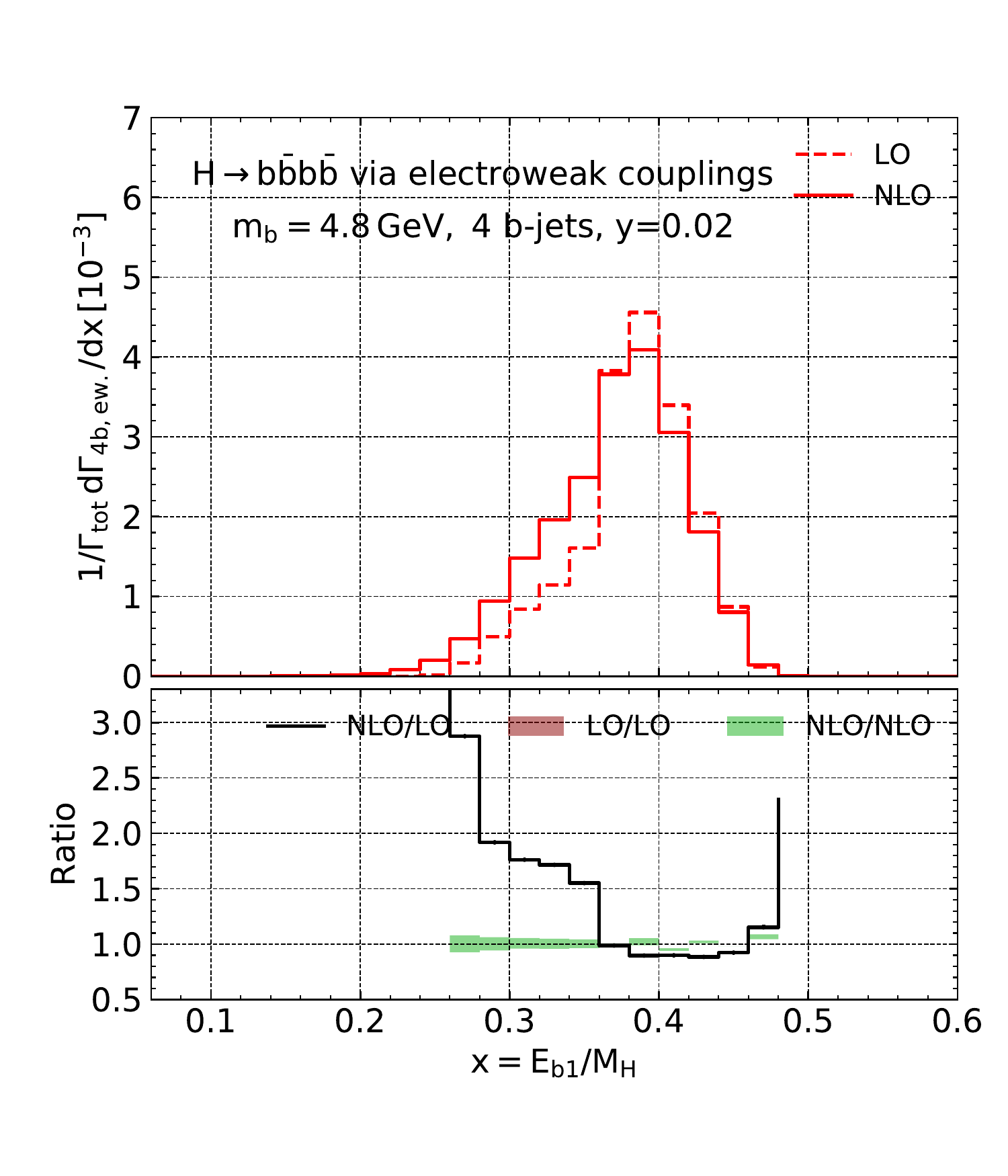}
\caption{
Differential decay branching ratio of the Higgs boson to four bottom quarks as a function
of the energy of the leading $b$-jet, via the Yukawa couplings (left plot) and
electroweak couplings (right plot), at both LO and NLO. 
\label{fig:eb1}}
\end{figure}

\begin{figure}[ht]
\centering
\includegraphics[width=0.47\textwidth]{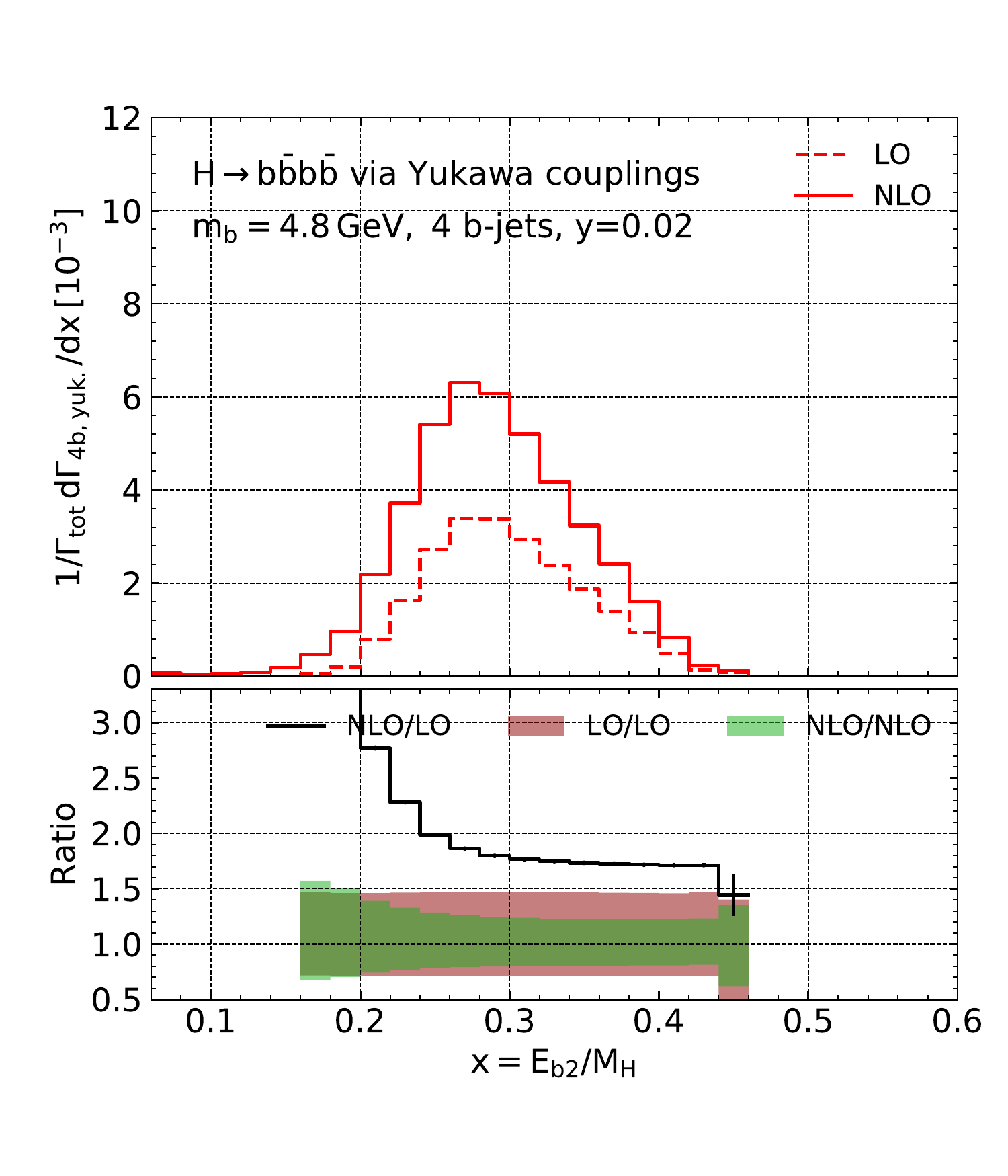}\hspace{0.1in}
\includegraphics[width=0.47\textwidth]{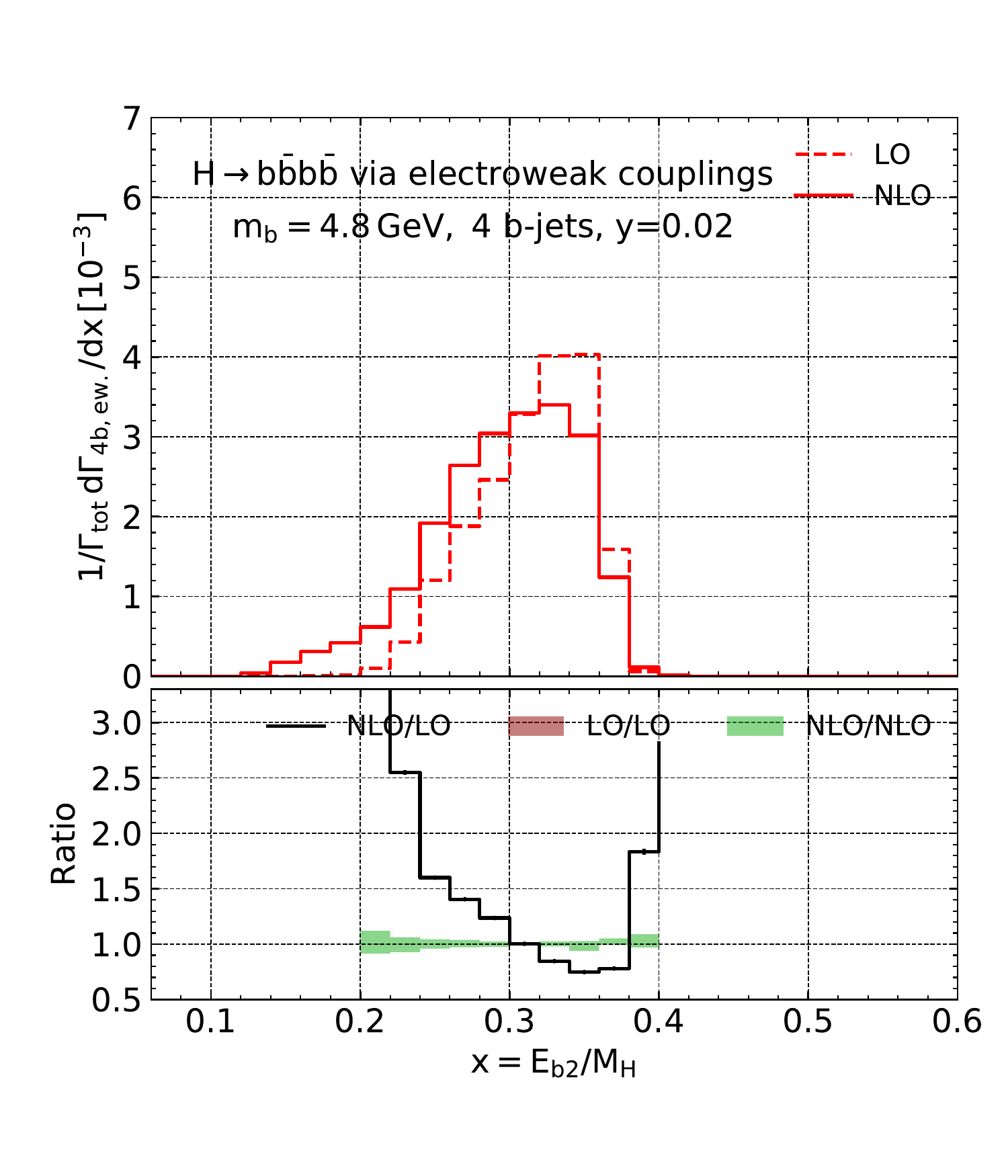}
\caption{
Similar to Fig.~\ref{fig:eb1} for distribution as a function of
of the energy of the sub-leading $b$-jet.
\label{fig:eb2}}
\end{figure}

\begin{figure}[ht]
\centering
\includegraphics[width=0.47\textwidth]{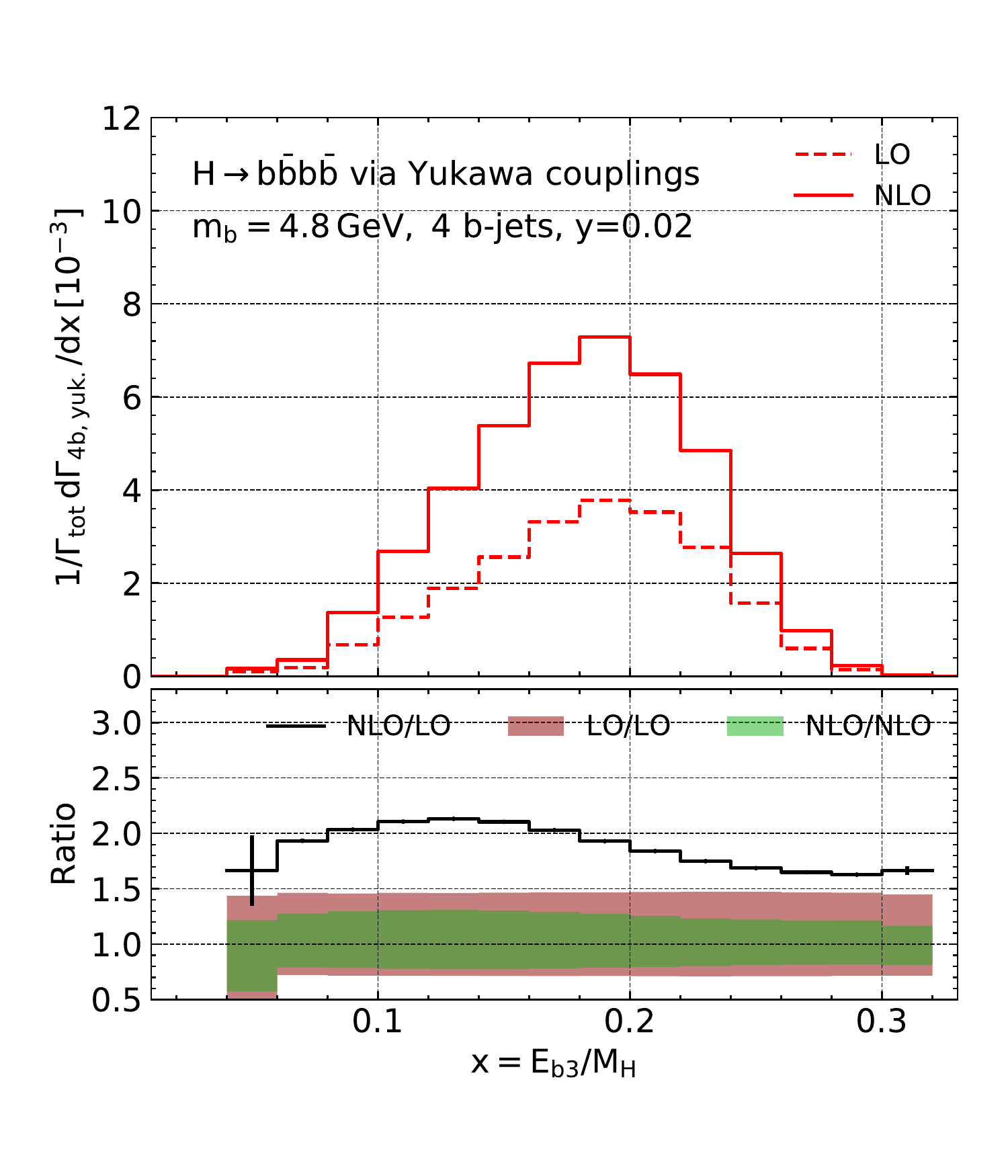}\hspace{0.1in}
\includegraphics[width=0.47\textwidth]{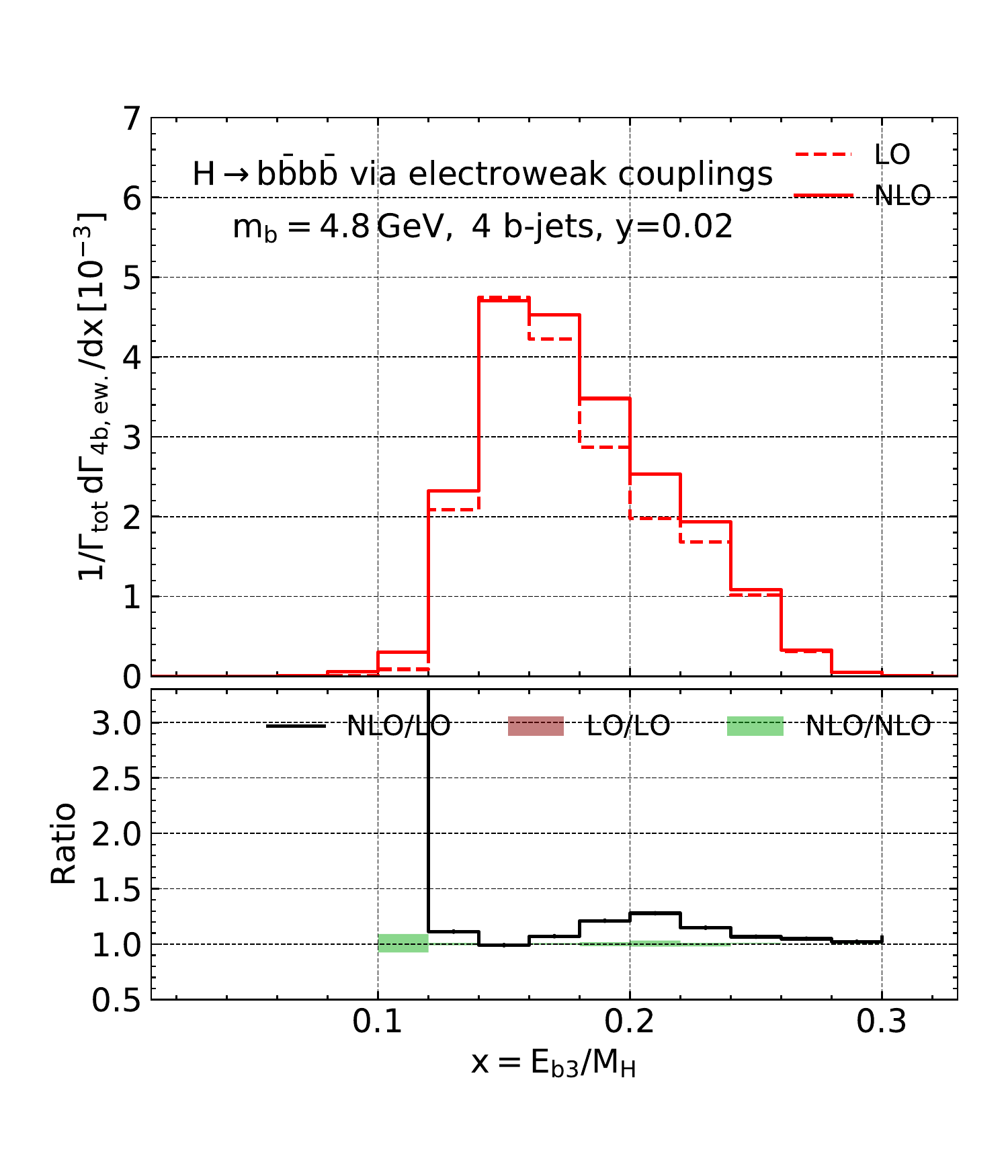}
\caption{
Similar to Fig.~\ref{fig:eb1} for distribution as a function of
of the energy of the third-leading $b$-jet.
\label{fig:eb3}}
\end{figure}

\begin{figure}[ht]
\centering
\includegraphics[width=0.47\textwidth]{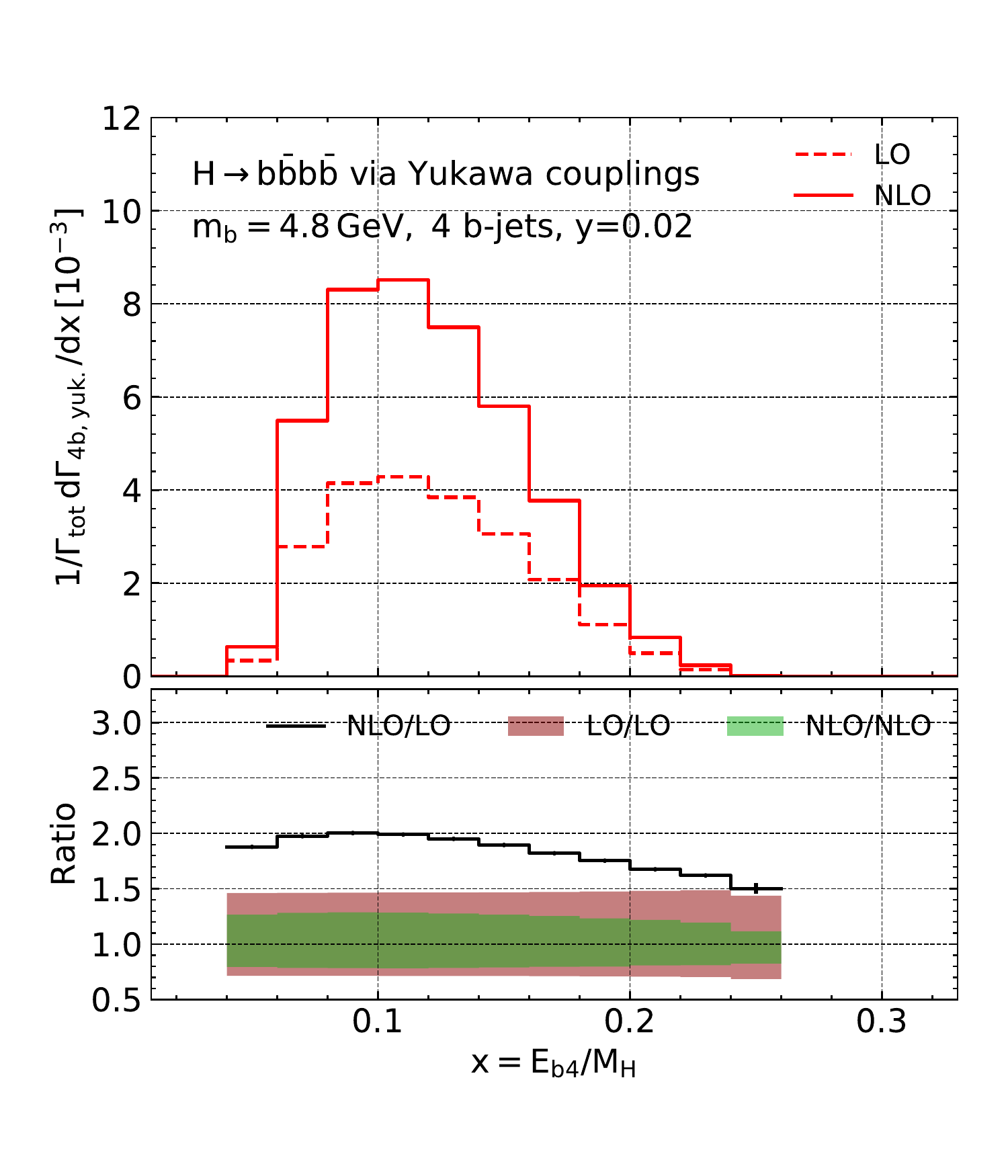}\hspace{0.1in}
\includegraphics[width=0.47\textwidth]{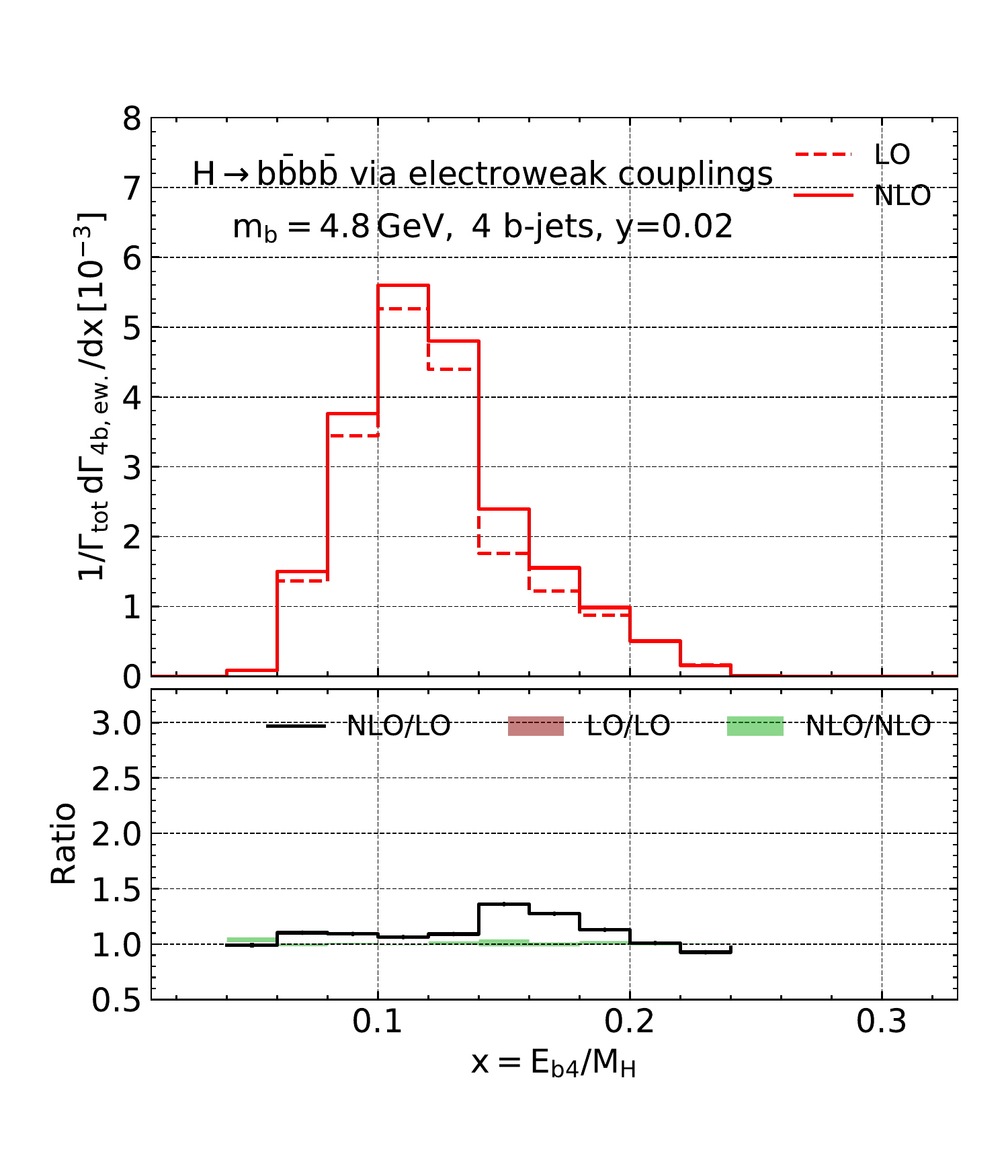}
\caption{
Similar to Fig.~\ref{fig:eb1} for distribution as a function of
of the energy of the softest $b$-jet.
\label{fig:eb4}}
\end{figure}

We show representative results for a choice of the jet resolution parameter
$y=0.02$ and $m_b=4.8$ GeV.
That corresponds to an angular separation of $\Delta \theta\sim 0.3$ for two
jets with the lower energy equals $M_H/4$.
We first plot the energy distributions of each individual jet in
Figs.~\ref{fig:eb1}-\ref{fig:eb4}.
Energies are always normalized to the mass of the Higgs boson.
We show the results for decay via Yukawa couplings and electroweak couplings
in parallel with upper panel gives the differential decay branching ratio and
lower inset gives ratio of NLO predictions to LO ones and their scale variations
respectively.
The distributions from two decay channels show clearly differences.
For instance, the (sub-)leading jet peaks at slightly lower(higher) values
for decay via electroweak couplings.
The energy spectrum of the third and fourth jets are broader for decay
via Yukawa couplings.
Beside of the normalization, the QCD corrections induce changes on shape
of the distributions.
The spectrum is generally pushed to lower energy region due to the recoiled
gluon in QCD real corrections, except for energy of the third and fourth
jets in decay via electroweak couplings, where enhancements are found right
after the peak region.  

\begin{figure}[ht]
\centering
\includegraphics[width=0.47\textwidth]{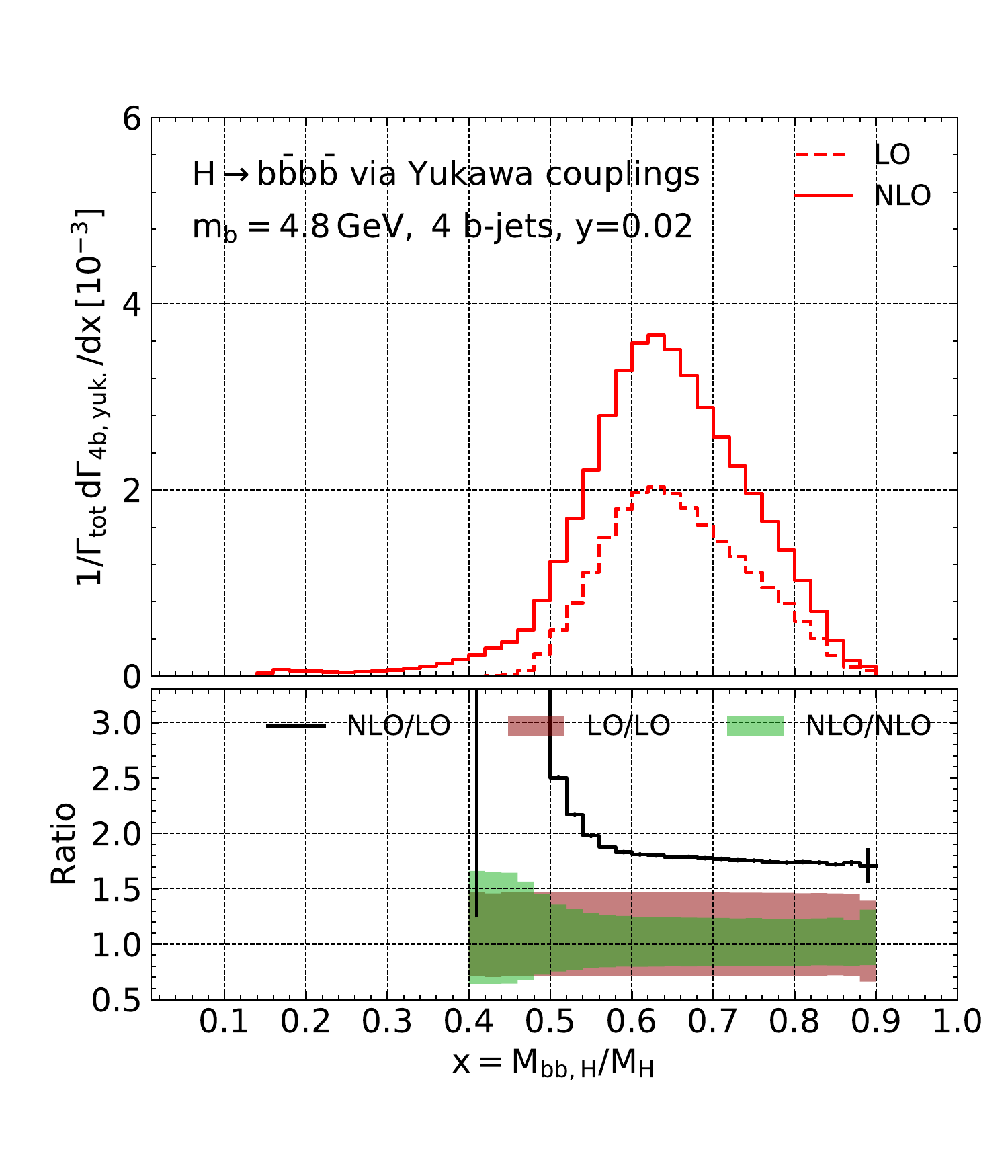}\hspace{0.1in}
\includegraphics[width=0.47\textwidth]{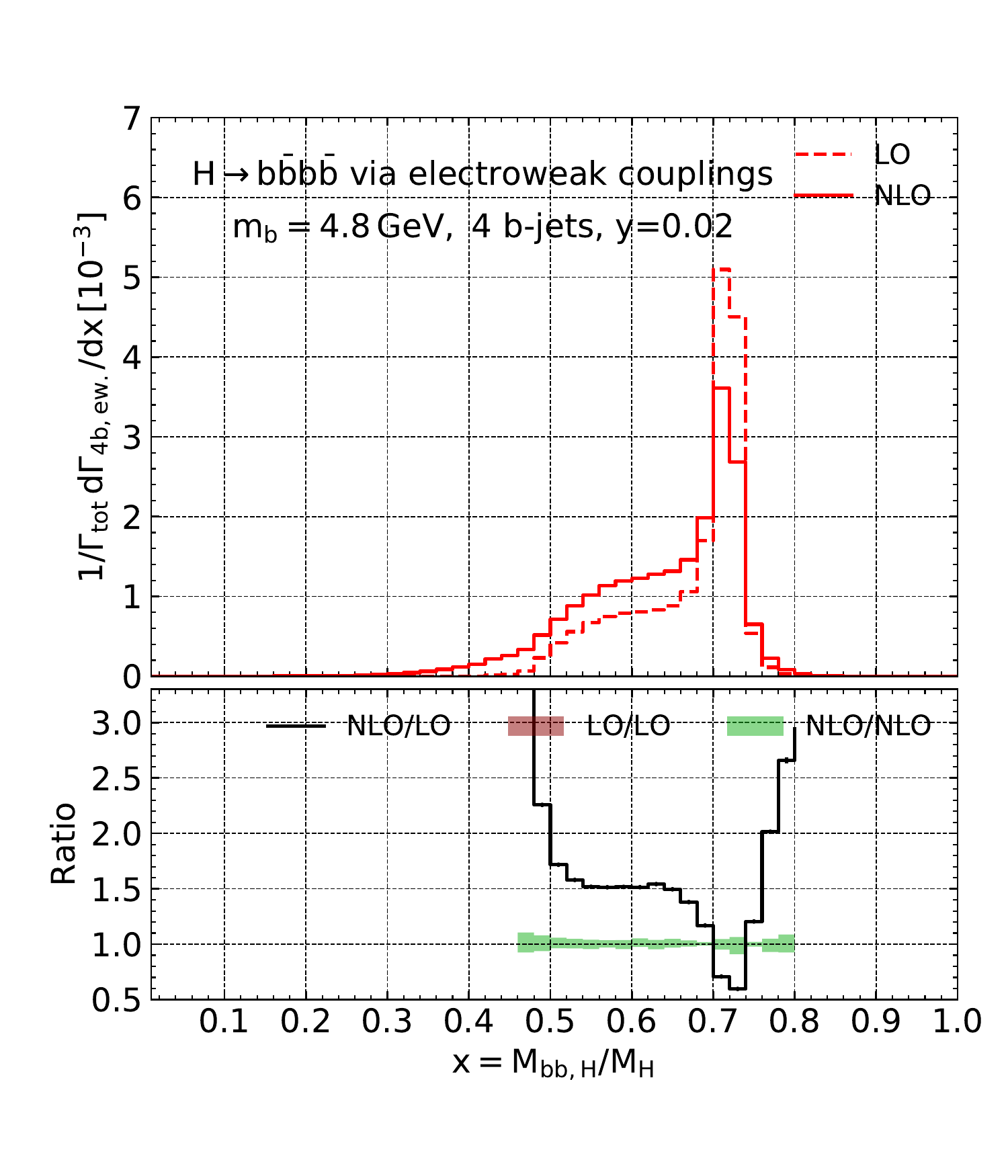}
\caption{
Similar to Fig.~\ref{fig:eb1} for distribution as a function of
of the highest invariant mass of all $b$-jet pairs.
\label{fig:mbbh}}
\end{figure}

\begin{figure}[ht]
\centering
\includegraphics[width=0.47\textwidth]{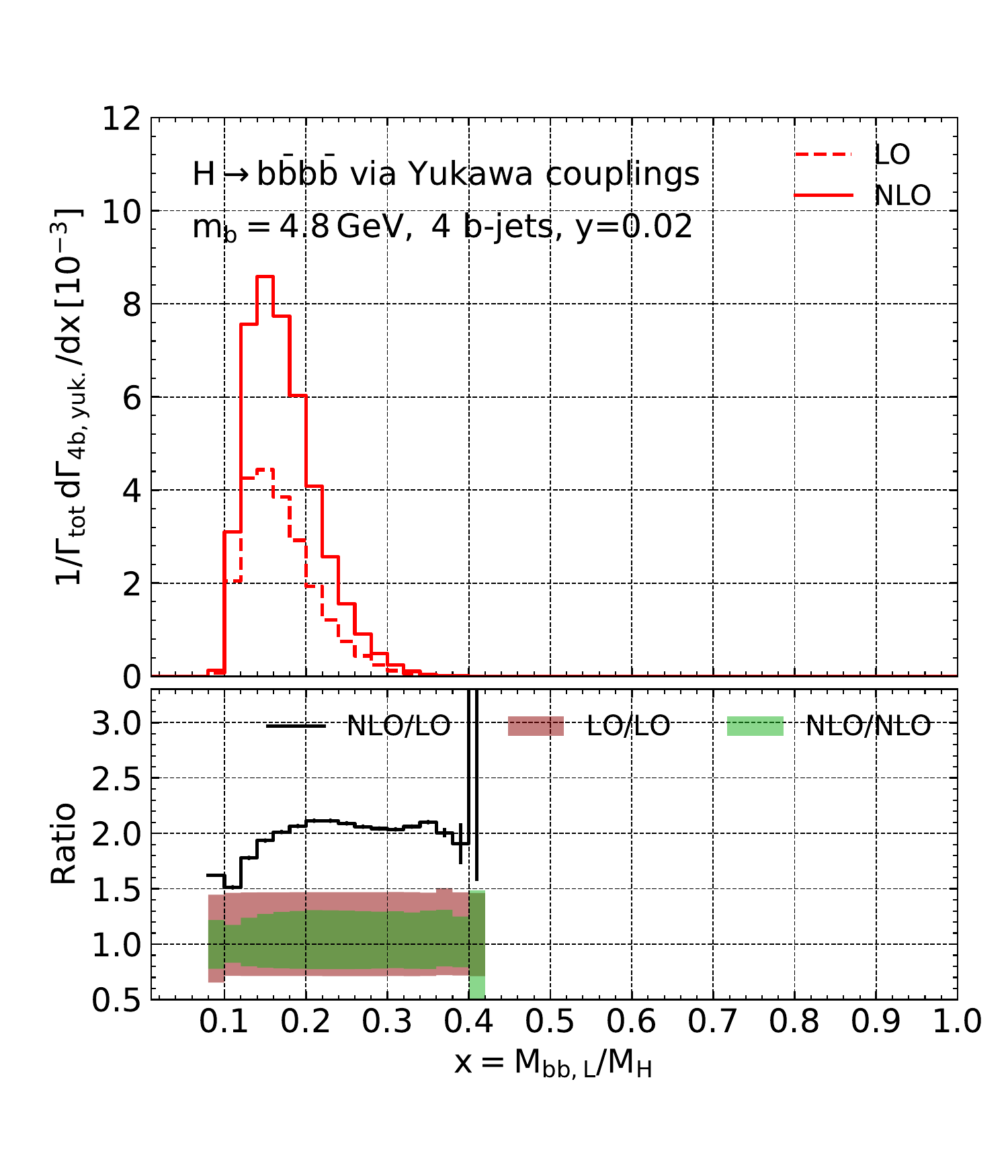}\hspace{0.1in}
\includegraphics[width=0.47\textwidth]{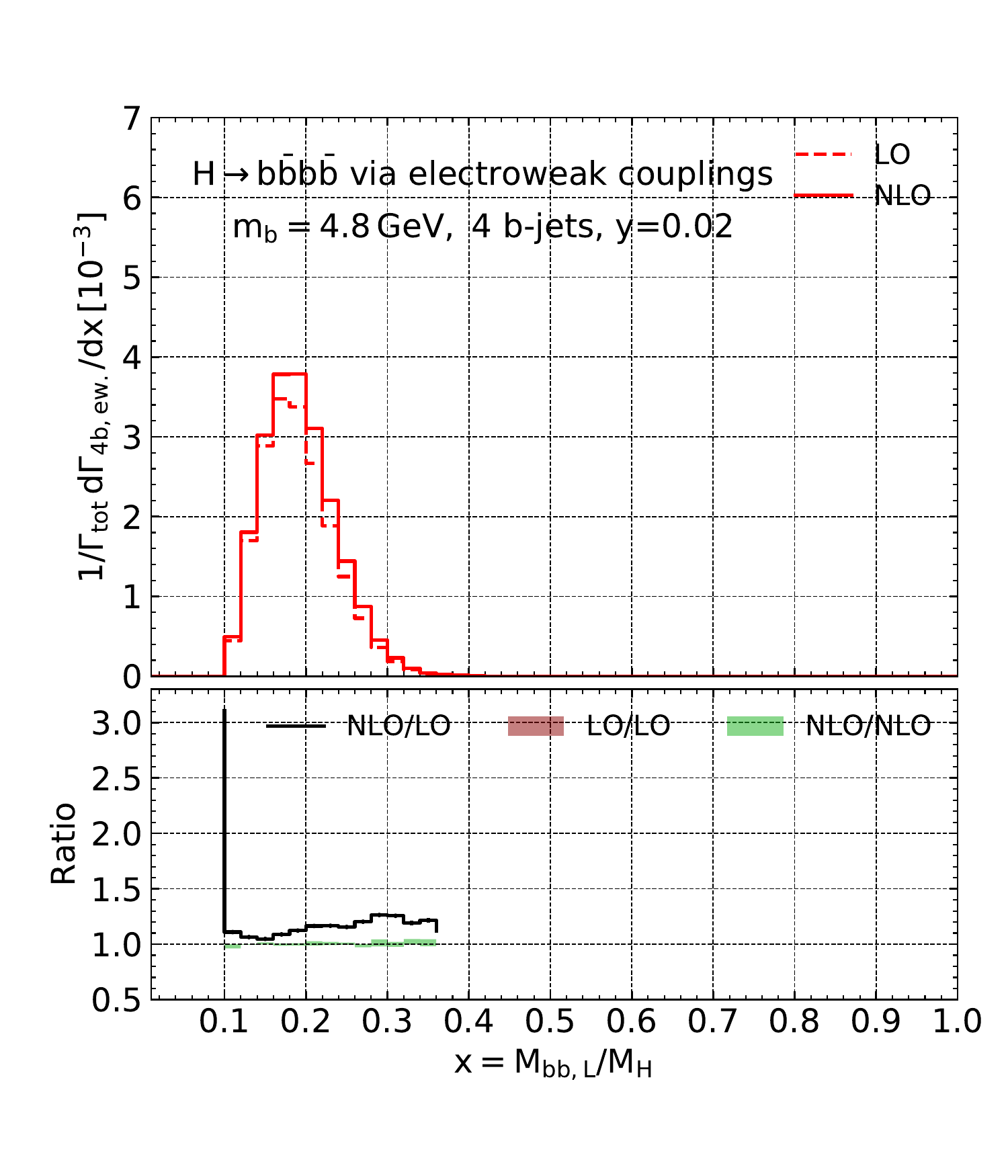}
\caption{
Similar to Fig.~\ref{fig:eb1} for distribution as a function of
the lowest invariant mass of all $b$-jet pairs.
\label{fig:mbbl}}
\end{figure}

\begin{figure}[ht]
\centering
\includegraphics[width=0.47\textwidth]{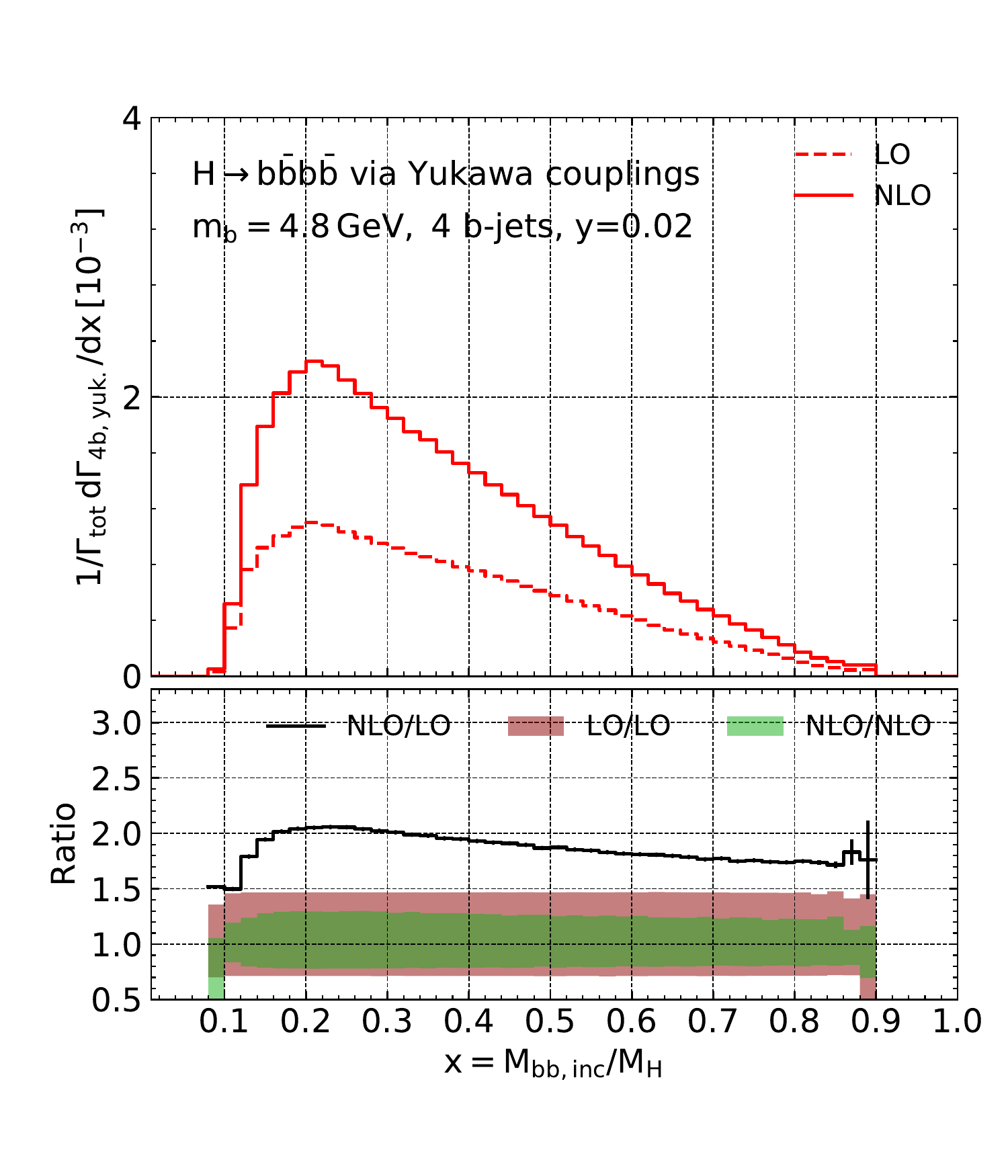}\hspace{0.1in}
\includegraphics[width=0.47\textwidth]{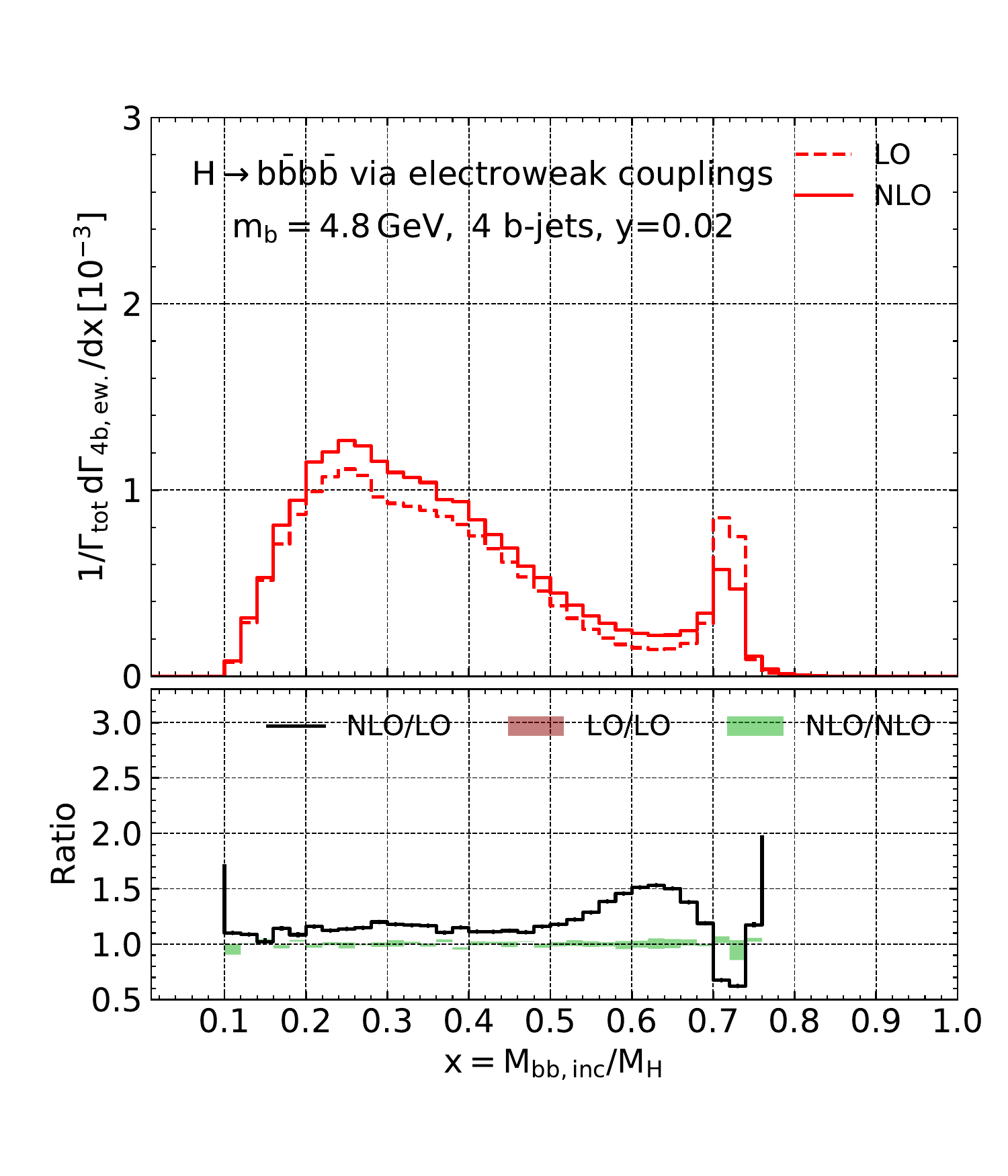}
\caption{
Similar to Fig.~\ref{fig:eb1} for distribution as a function of
the inclusive invariant mass of all $b$-jet pairs.
\label{fig:mbbi}}
\end{figure}

\begin{figure}[ht]
\centering
\includegraphics[width=0.47\textwidth]{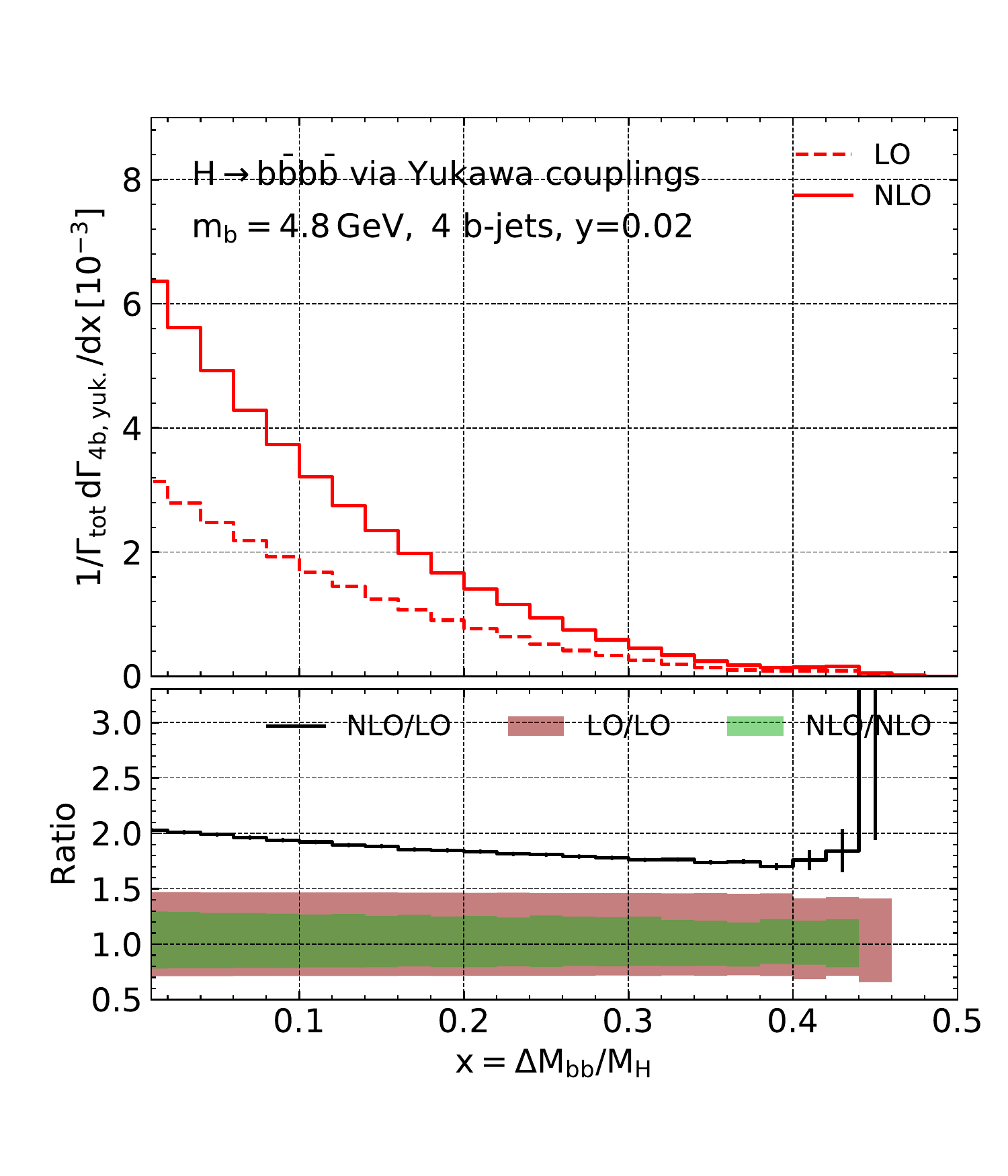}\hspace{0.1in}
\includegraphics[width=0.47\textwidth]{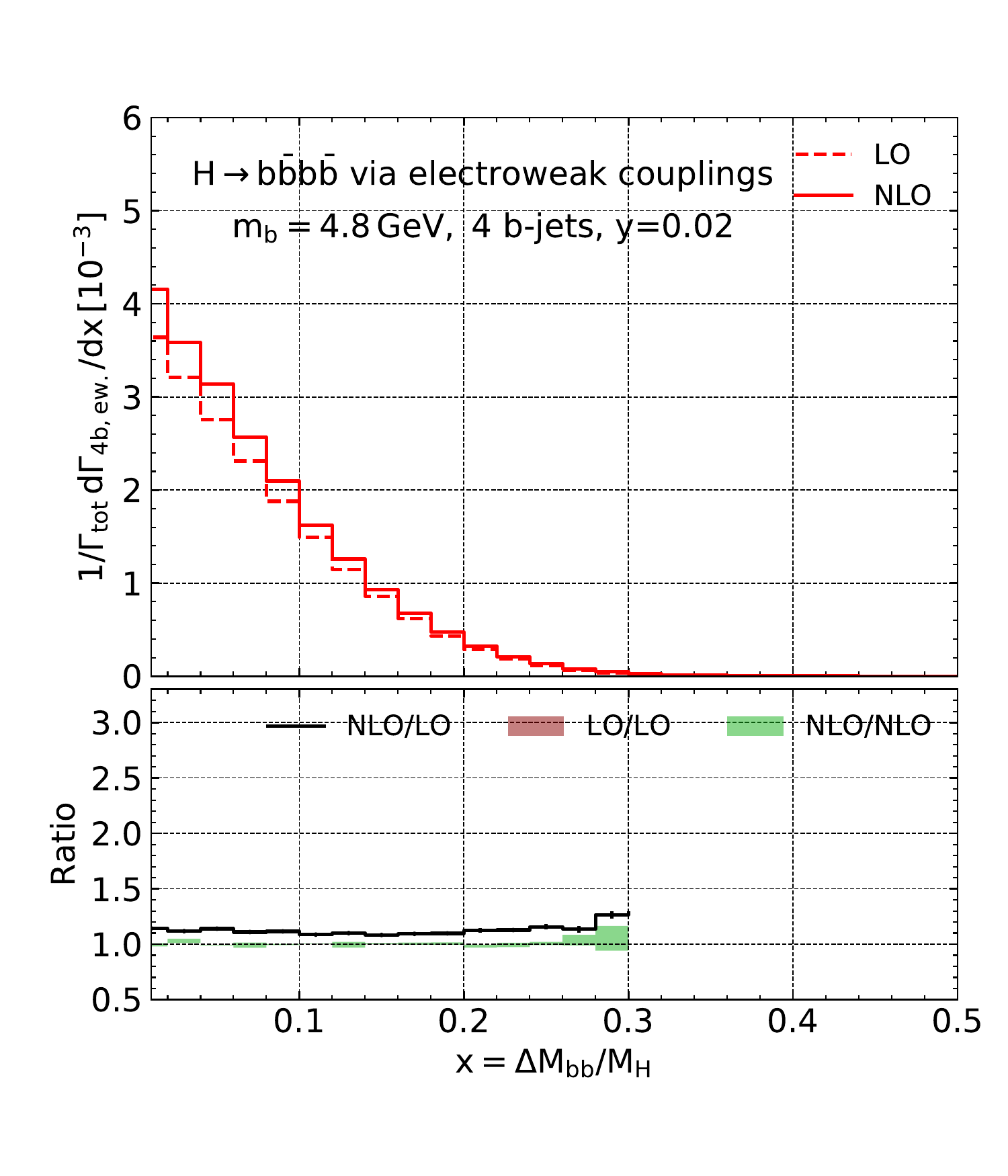}
\caption{
Similar to Fig.~\ref{fig:eb1} for distribution as a function of
the invariant mass asymmetry of all $b$-jet pairs.
\label{fig:dmbb}}
\end{figure}

We now move to distributions of invariant mass of jet pairs.
In Fig.~\ref{fig:mbbh} we plot distribution of the highest invariant
mass among all jet pairs in the event, $M_{b\bar b, H}$.
The decay via Yukawa couplings show a broad peak around 0.6$M_H$.
On another hand decay via electroweak couplings has a narrow peak
close to 0.7$M_H$ due to the dominant contributions from onshell
$Z$ boson.
Concerning shape of the distributions, the QCD corrections shift
the spectrum to lower mass region for decay via Yukawa couplings.
Both sides off the narrow peak are enhanced for decay via electroweak
couplings because of the QCD real radiations.
For the distribution of the lowest invariant mass $M_{b\bar b,L}$ in
Fig.~\ref{fig:mbbl},
the decay via Yukawa couplings shows a sharp peak right after the
mass threshold.
In both decay channels the QCD corrections turn to induce a change of spectrum
towards high mass regions.
We plot the inclusive invariant mass distribution of jet pairs in
Fig.~\ref{fig:mbbi}.
They show a much broader spectrum as expected and peak around
0.2$M_H$.
The invariant mass peak at $Z$ boson mass in decay via electroweak
couplings are diluted for the inclusive mass distribution.
The QCD corrections sharpen the peak slightly in the case of decay
via Yukawa couplings.
Finally in Fig.~\ref{fig:dmbb} we show distributions of the invariant
mass asymmetry.
Both channels can have rather large asymmetry and show similar shapes.
The QCD corrections are stable crossing the full kinematic range.

\begin{figure}[ht]
\centering
\includegraphics[width=0.47\textwidth]{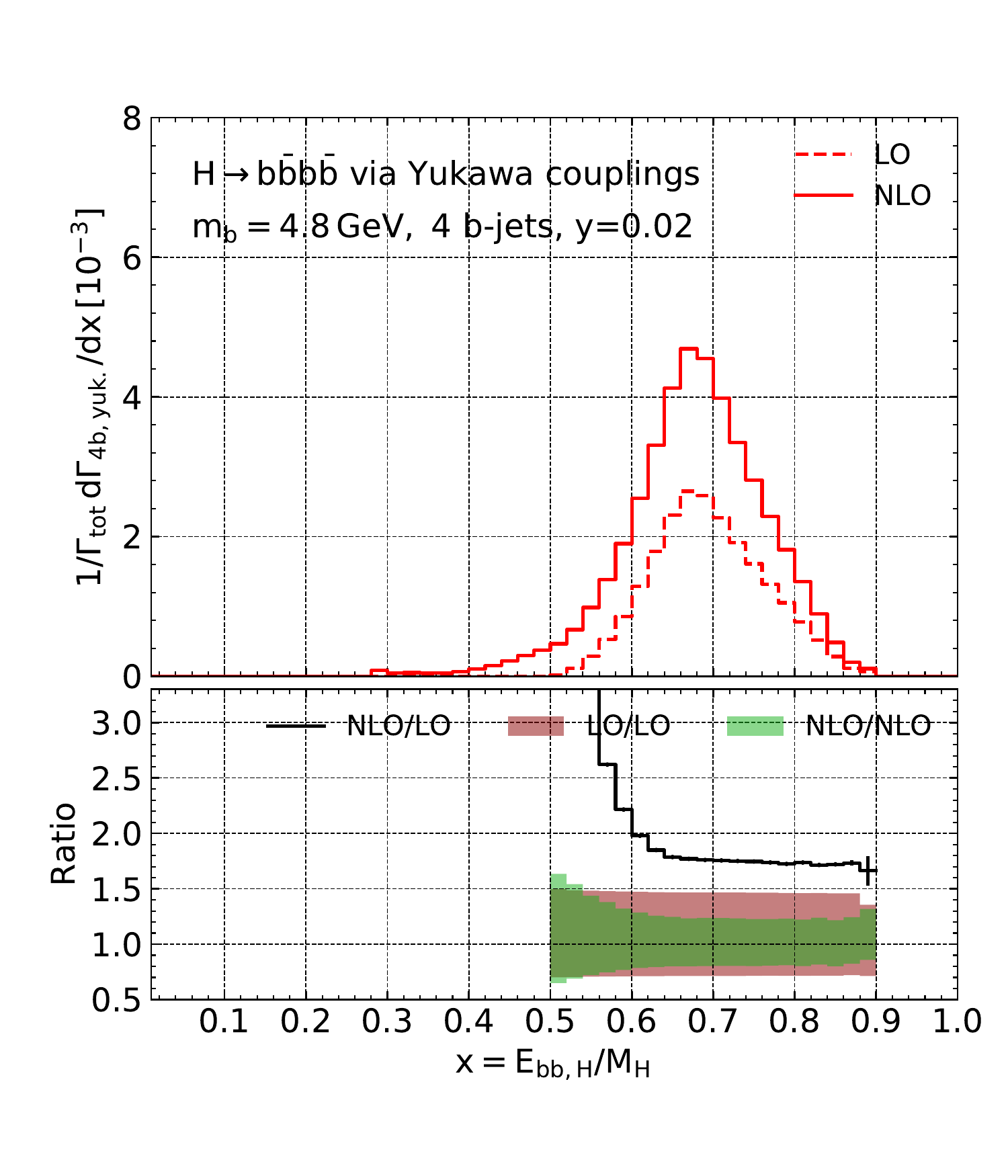}\hspace{0.1in}
\includegraphics[width=0.47\textwidth]{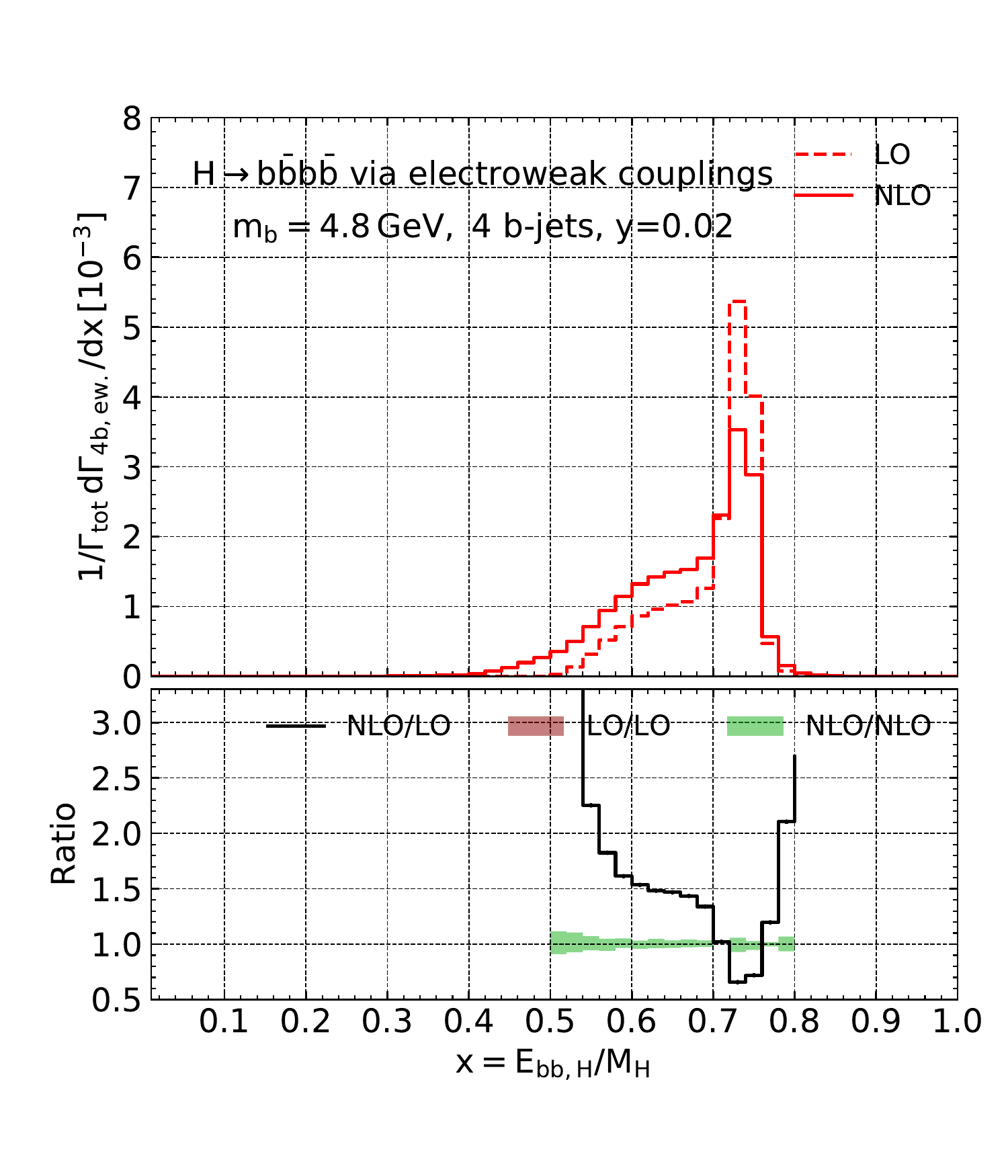}
\caption{
Similar to Fig.~\ref{fig:eb1} for distribution as a function of
the highest energy of all $b$-jet pairs.
\label{fig:ebbh}}
\end{figure}

\begin{figure}[ht]
\centering
\includegraphics[width=0.47\textwidth]{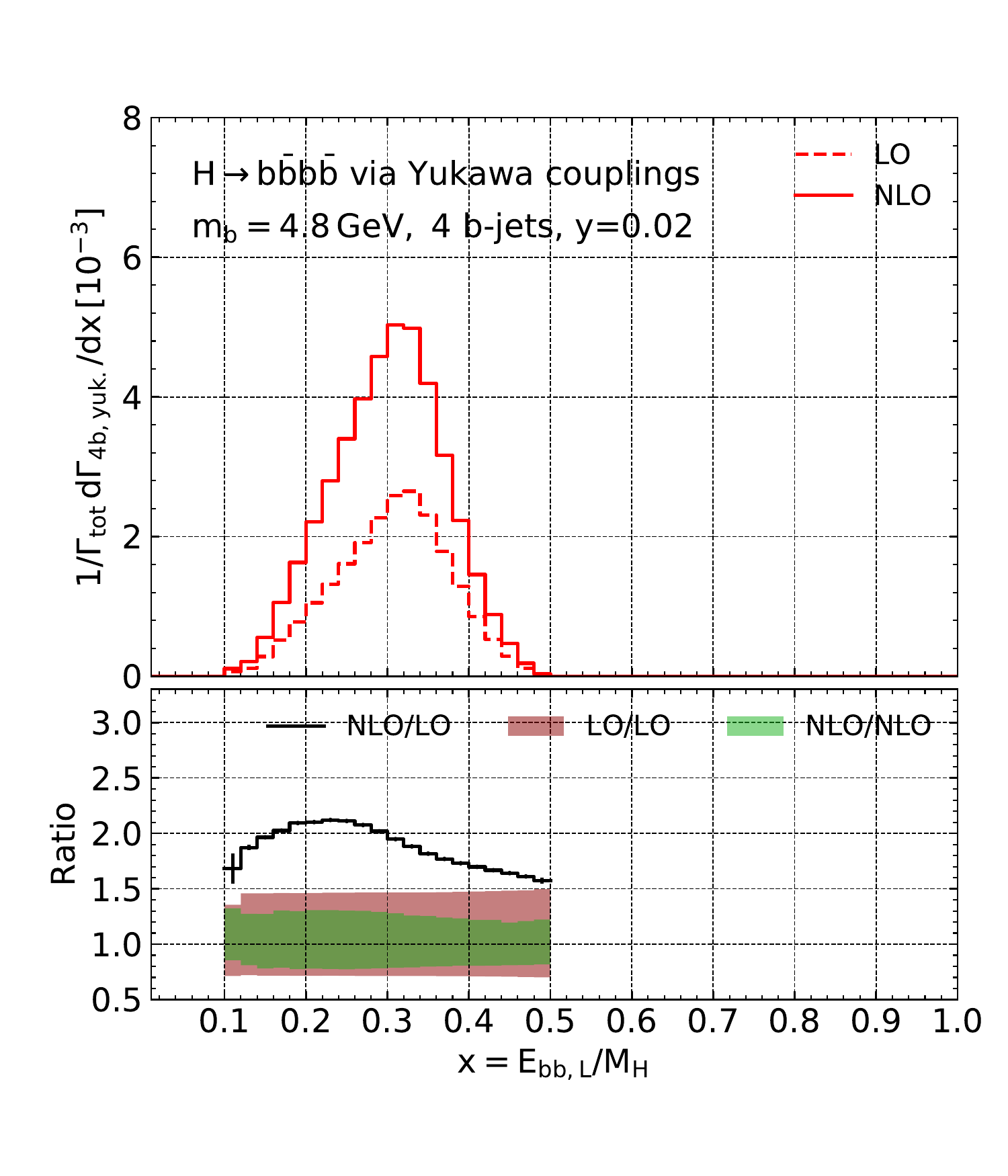}\hspace{0.1in}
\includegraphics[width=0.47\textwidth]{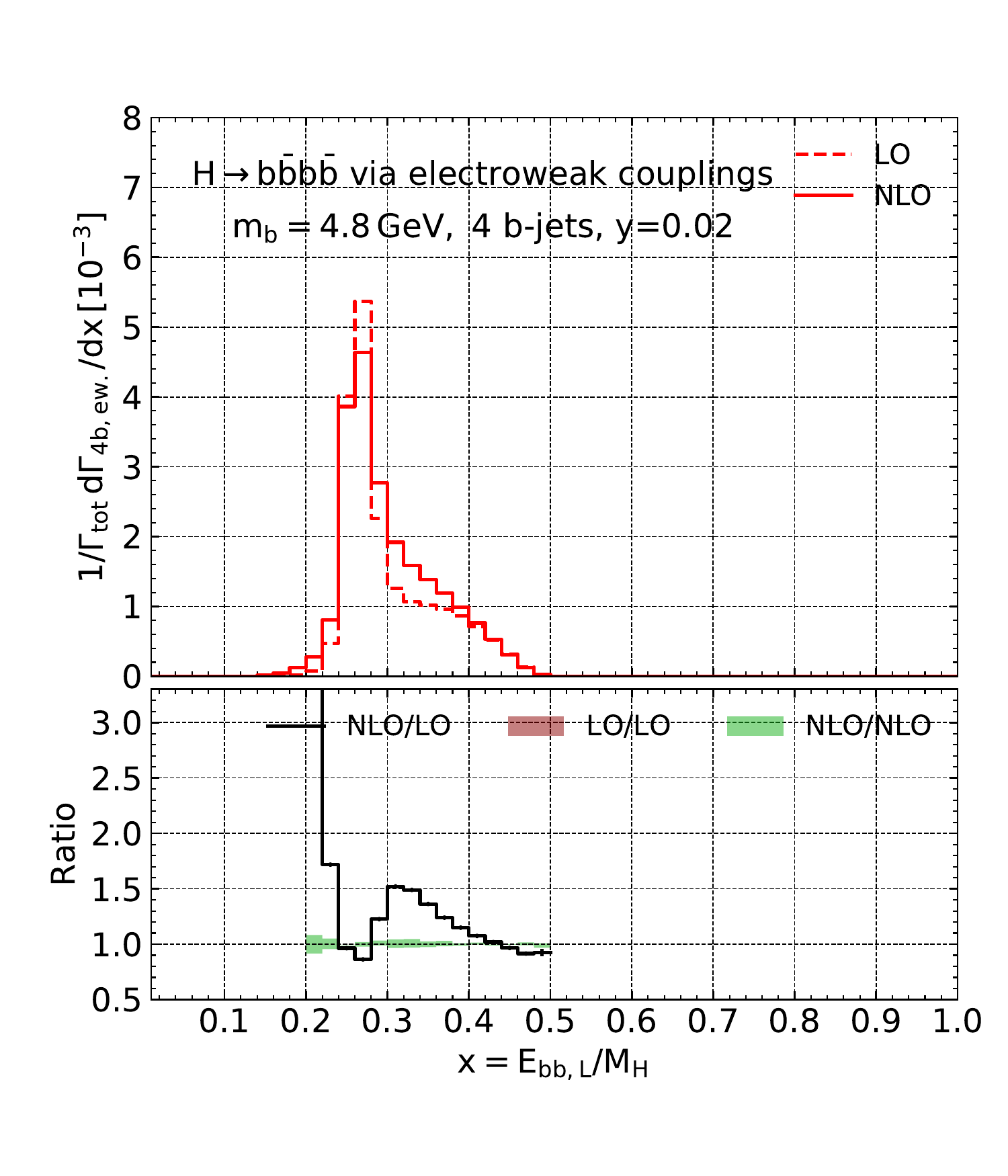}
\caption{
Similar to Fig.~\ref{fig:eb1} for distribution as a function of
the lowest energy of all $b$-jet pairs.
\label{fig:ebbl}}
\end{figure}

\begin{figure}[ht]
\centering
\includegraphics[width=0.47\textwidth]{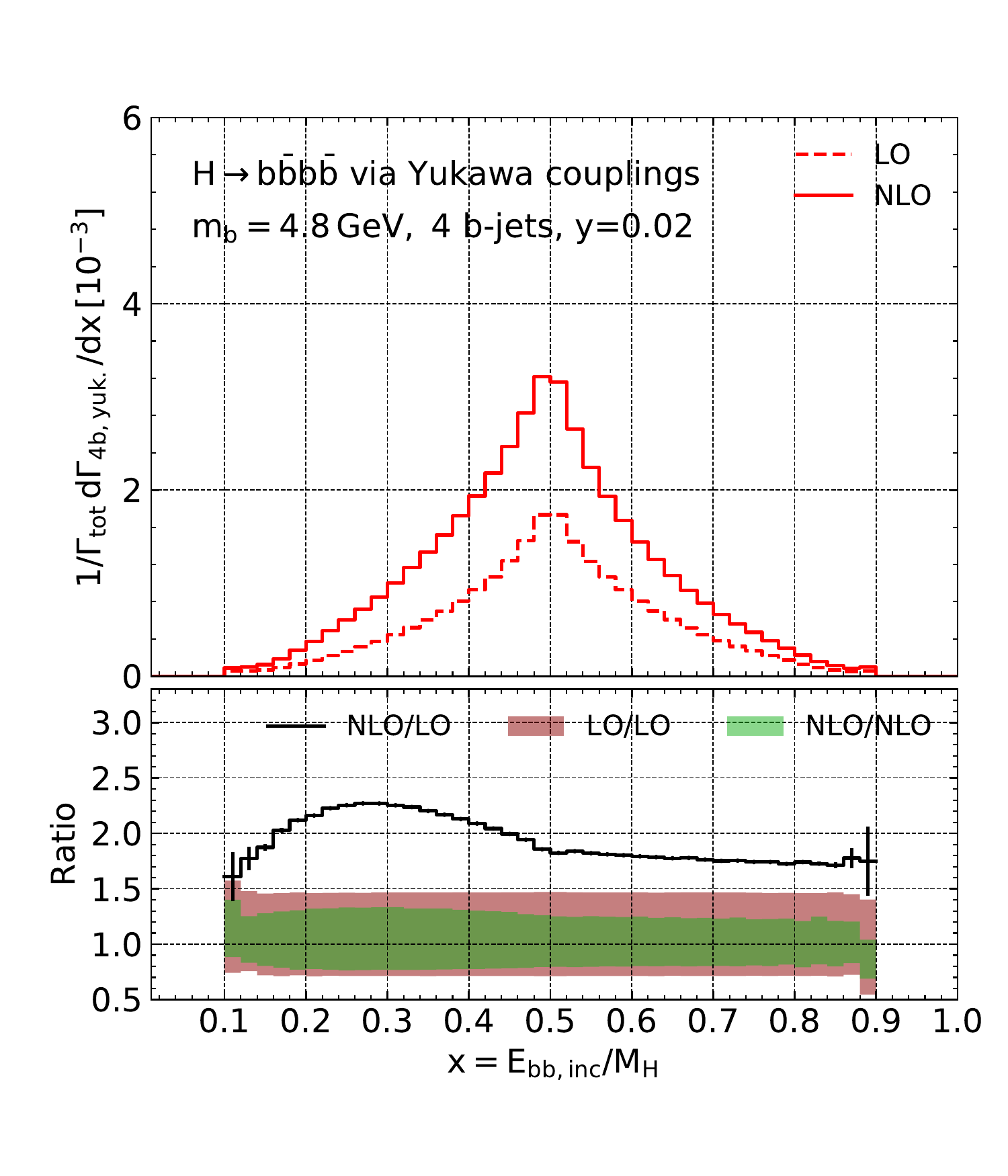}\hspace{0.1in}
\includegraphics[width=0.47\textwidth]{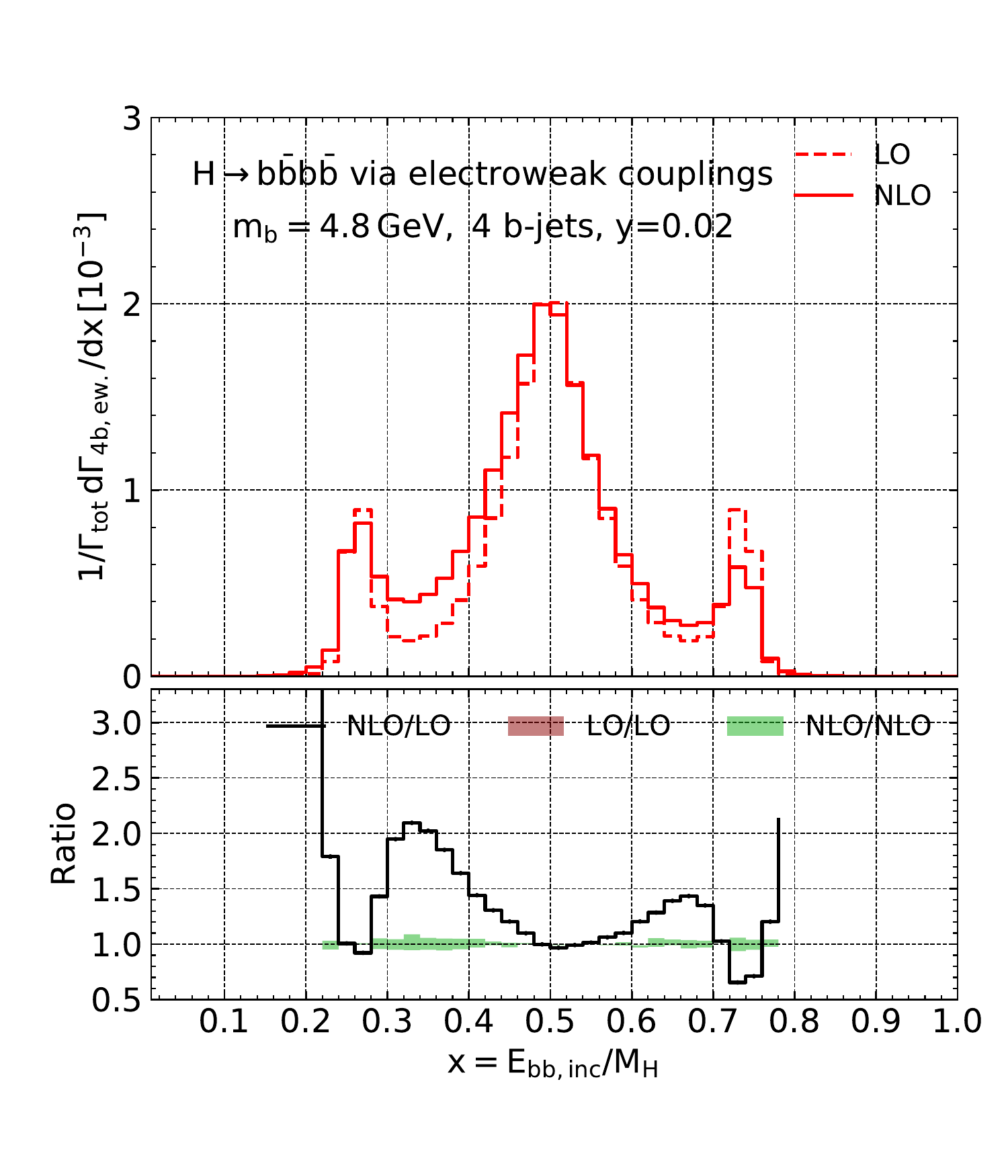}
\caption{
Similar to Fig.~\ref{fig:eb1} for distribution as a function of
the inclusive energy of all $b$-jet pairs.
\label{fig:ebbi}}
\end{figure}

\begin{figure}[ht]
\centering
\includegraphics[width=0.47\textwidth]{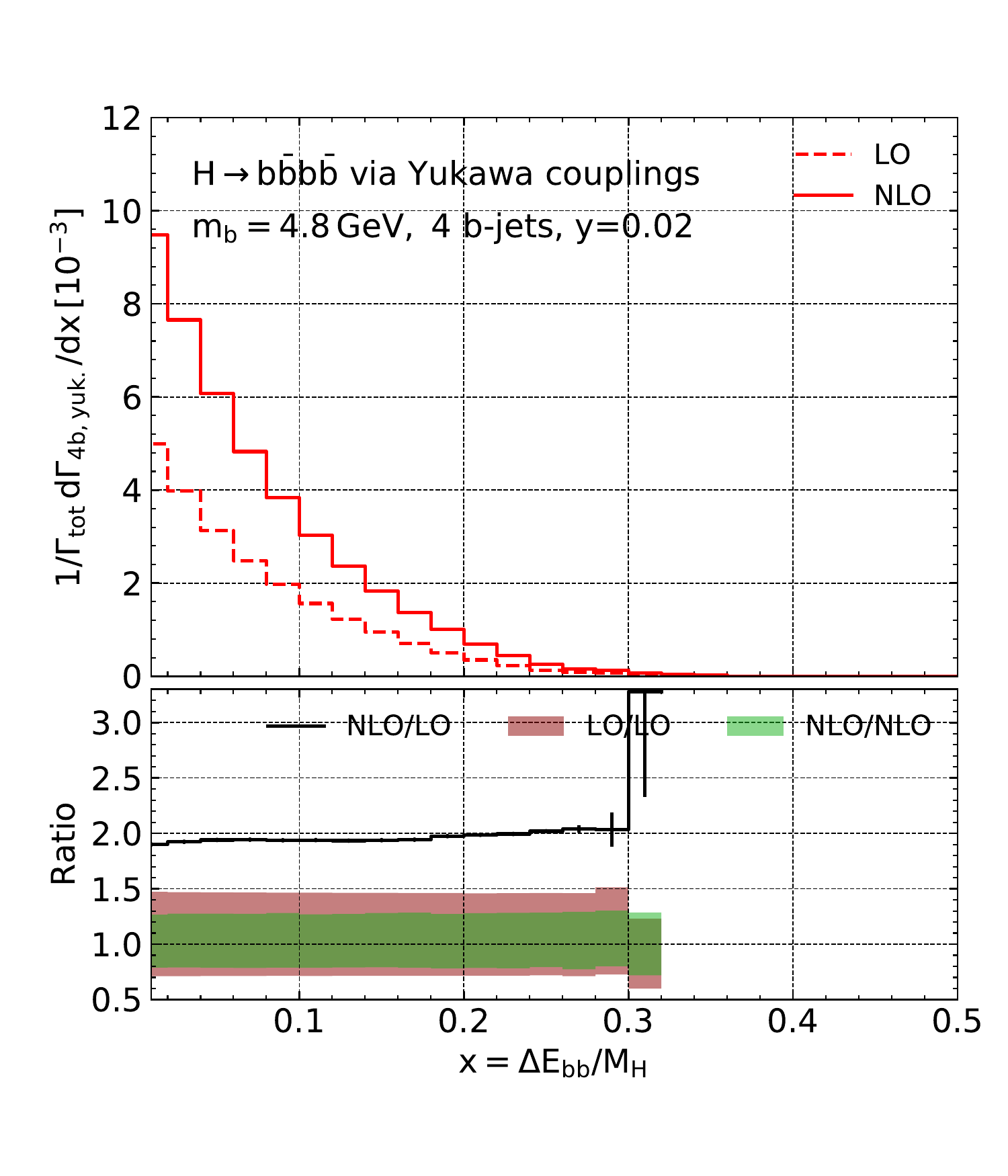}\hspace{0.1in}
\includegraphics[width=0.47\textwidth]{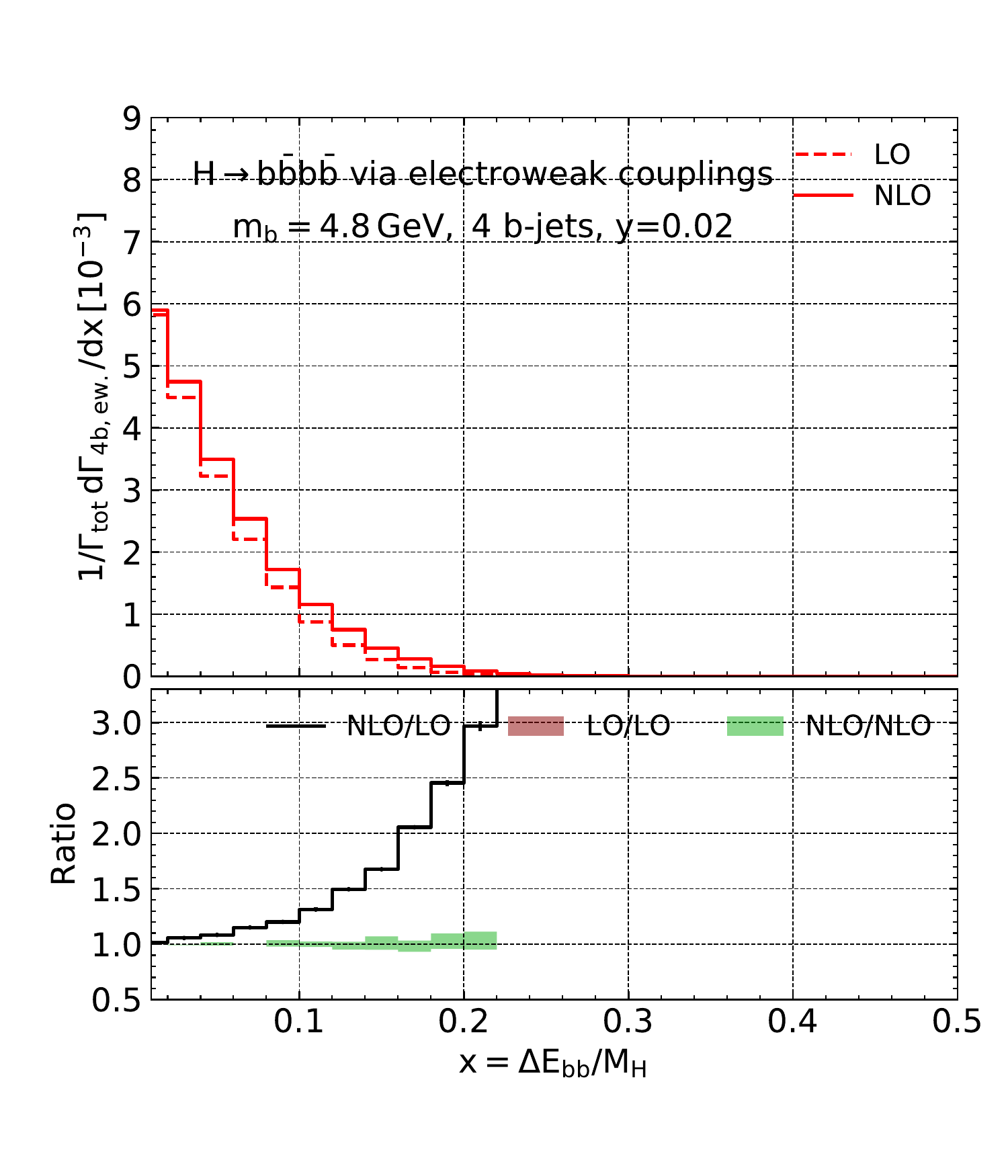}
\caption{
Similar to Fig.~\ref{fig:eb1} for distribution as a function of
the energy asymmetry of all $b$-jet pairs.
\label{fig:debb}}
\end{figure}

We present various distributions on energy of the jet pairs in
Figs.~\ref{fig:ebbh}-\ref{fig:debb}.
The distribution on the highest energy of all jet pairs in Fig.~\ref{fig:ebbh} tends to
be very similar to the case of highest invariant mass as discussed
in Fig.~\ref{fig:mbbh} since they are mostly from the same jet pair.
The impact of QCD corrections are also similar, namely push the spectrum
to lower energy region for the decay through Yukawa couplings and broaden
the $Z$ mass peak in the case of electroweak couplings.
For the distribution on the lowest energy of all jet pairs in Fig.~\ref{fig:ebbl},
at LO it is simply a reflection of the distribution on the highest energy
of all jet pairs since the two energies add up to the Higgs boson mass.
The QCD radiations change the shape of the distribution in a non-trivial way.
Due to the same reason the inclusive energy distribution of all jet pairs in
Fig.~\ref{fig:ebbi} are symmetric with respect to an energy of $M_H/2$ at LO.
Thus the decay via electroweak couplings exhibits a triple-peak structure.
For their shape the QCD corrections push the distribution to lower energy
region in general due to energies carried away by gluon radiations.
Energy of jet pairs can also show large asymmetry as presented in Fig.~\ref{fig:debb}.
The QCD corrections are almost constant for decay via Yukawa couplings and
are largely enhanced towards the tail of the distribution for decay via
electroweak couplings.   

\begin{figure}[ht]
\centering
\includegraphics[width=0.4\textwidth]{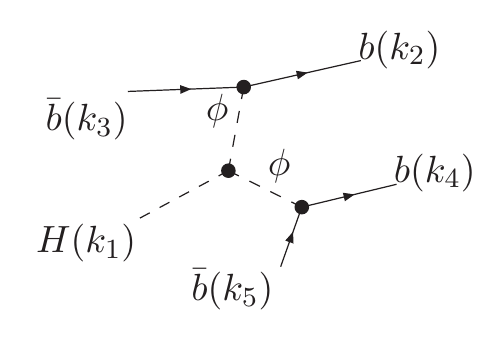}
\caption{
Feynman diagram at leading order for exotic decay of the Higgs boson to four bottom
quarks via two light scalars.
\label{fig:ff3}}
\end{figure}

\begin{figure}[ht]
\centering
\includegraphics[width=0.47\textwidth]{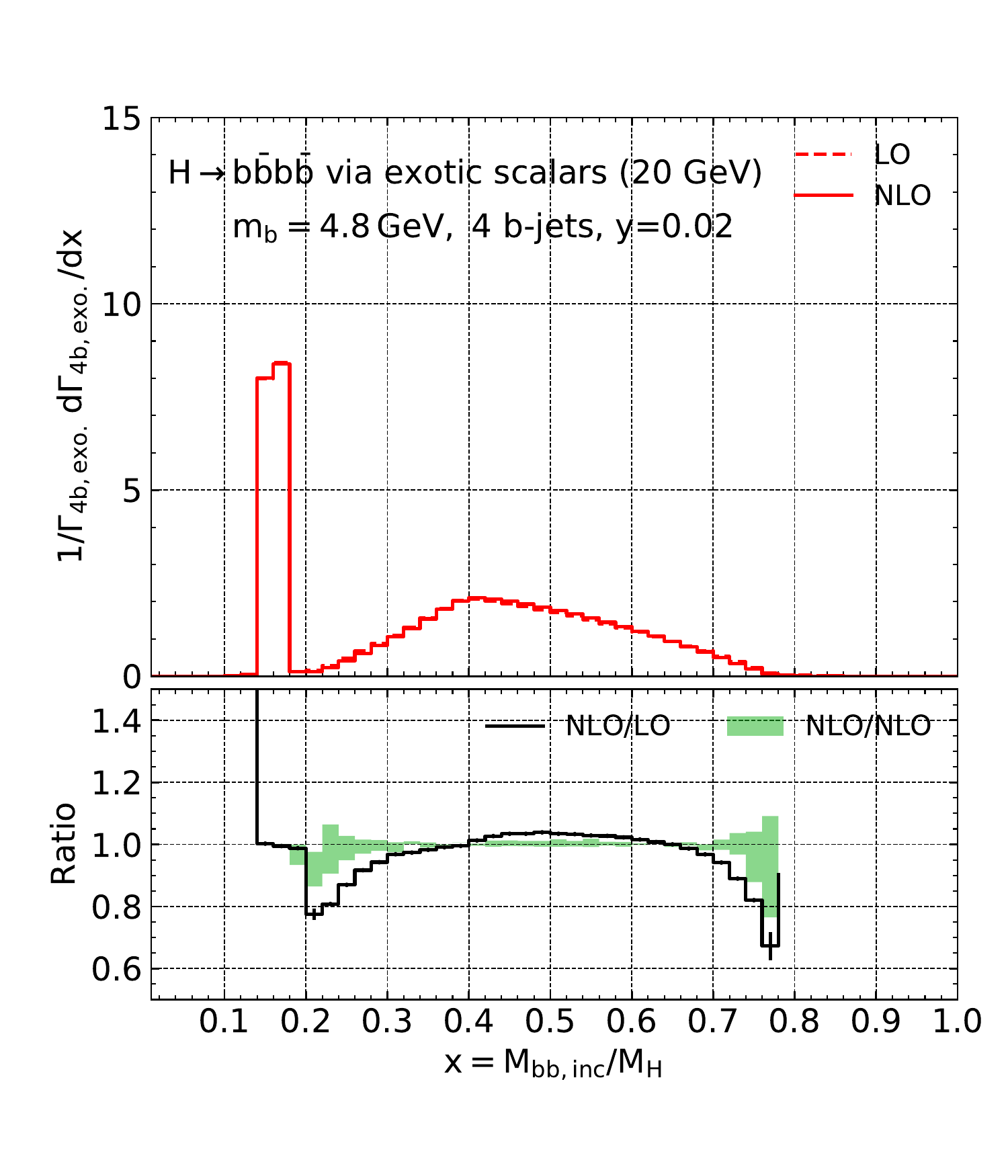}\hspace{0.1in}
\includegraphics[width=0.47\textwidth]{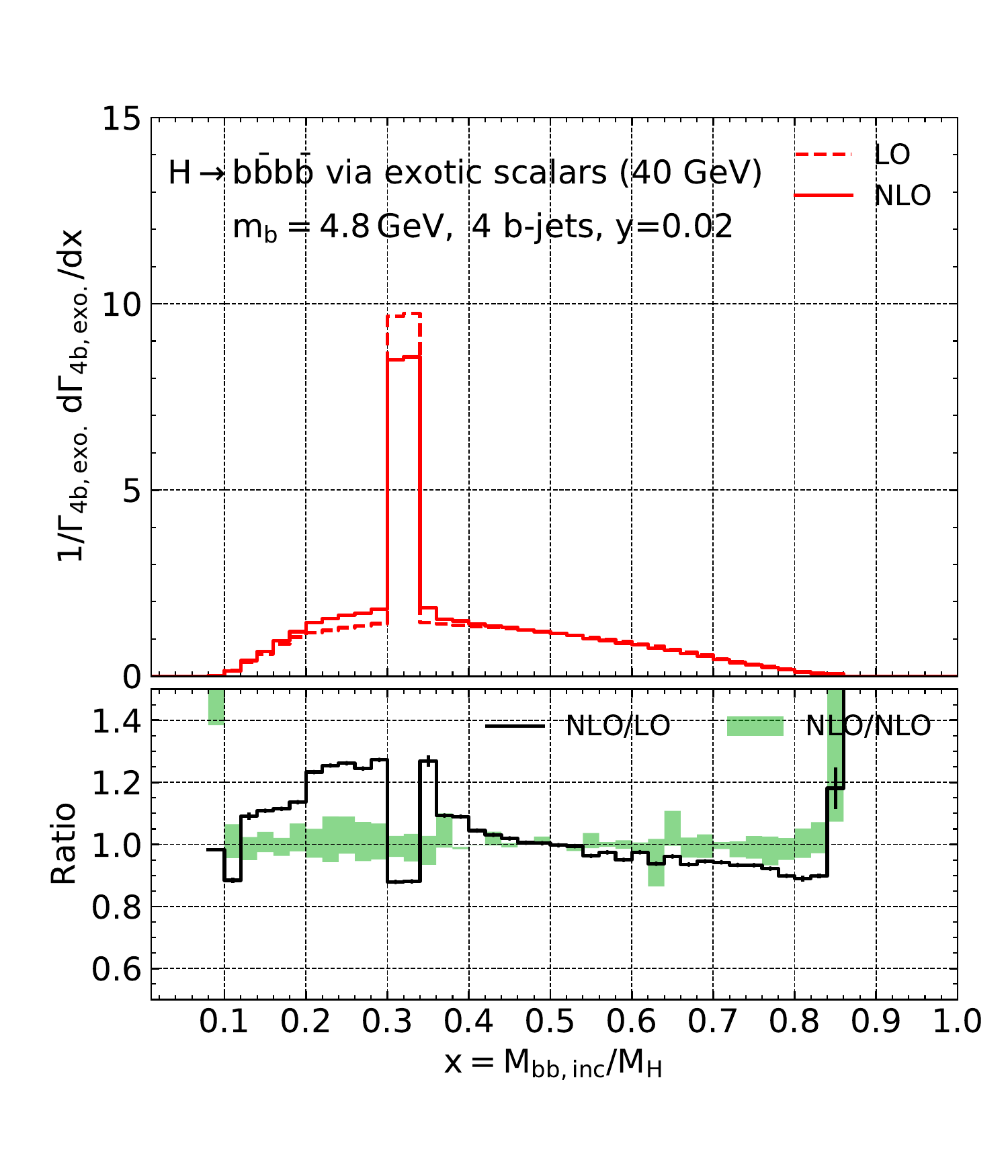}
\caption{
Normalized distribution of the Higgs boson decay to four bottom quarks as a function
of the inclusive invariant mass of all $b$-jet pairs, via new light scalars with a
mass of 20 GeV (left plot) and 40 GeV (right plot), at both LO and NLO. 
\label{fig:Ambbi}}
\end{figure}

\begin{figure}[ht]
\centering
\includegraphics[width=0.47\textwidth]{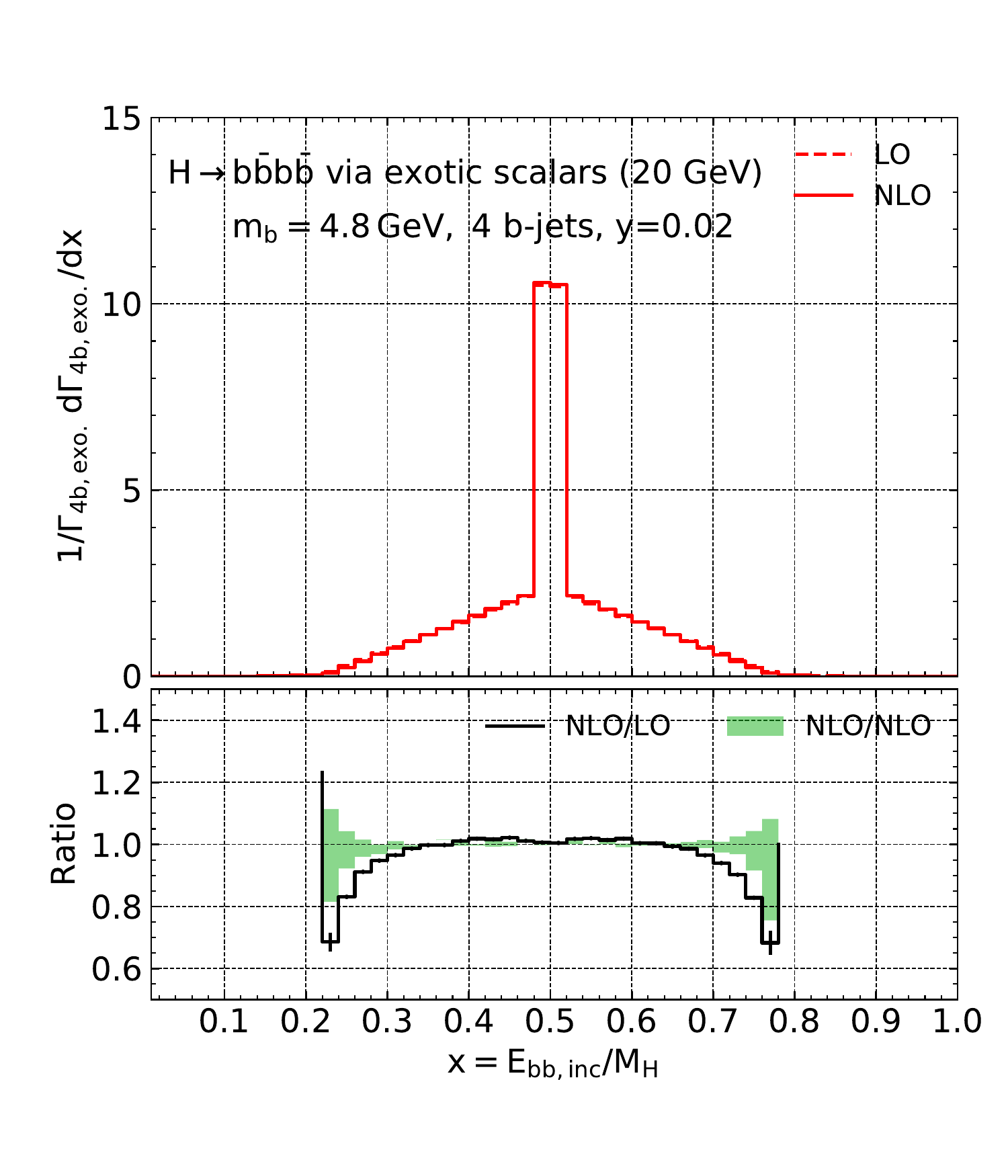}\hspace{0.1in}
\includegraphics[width=0.47\textwidth]{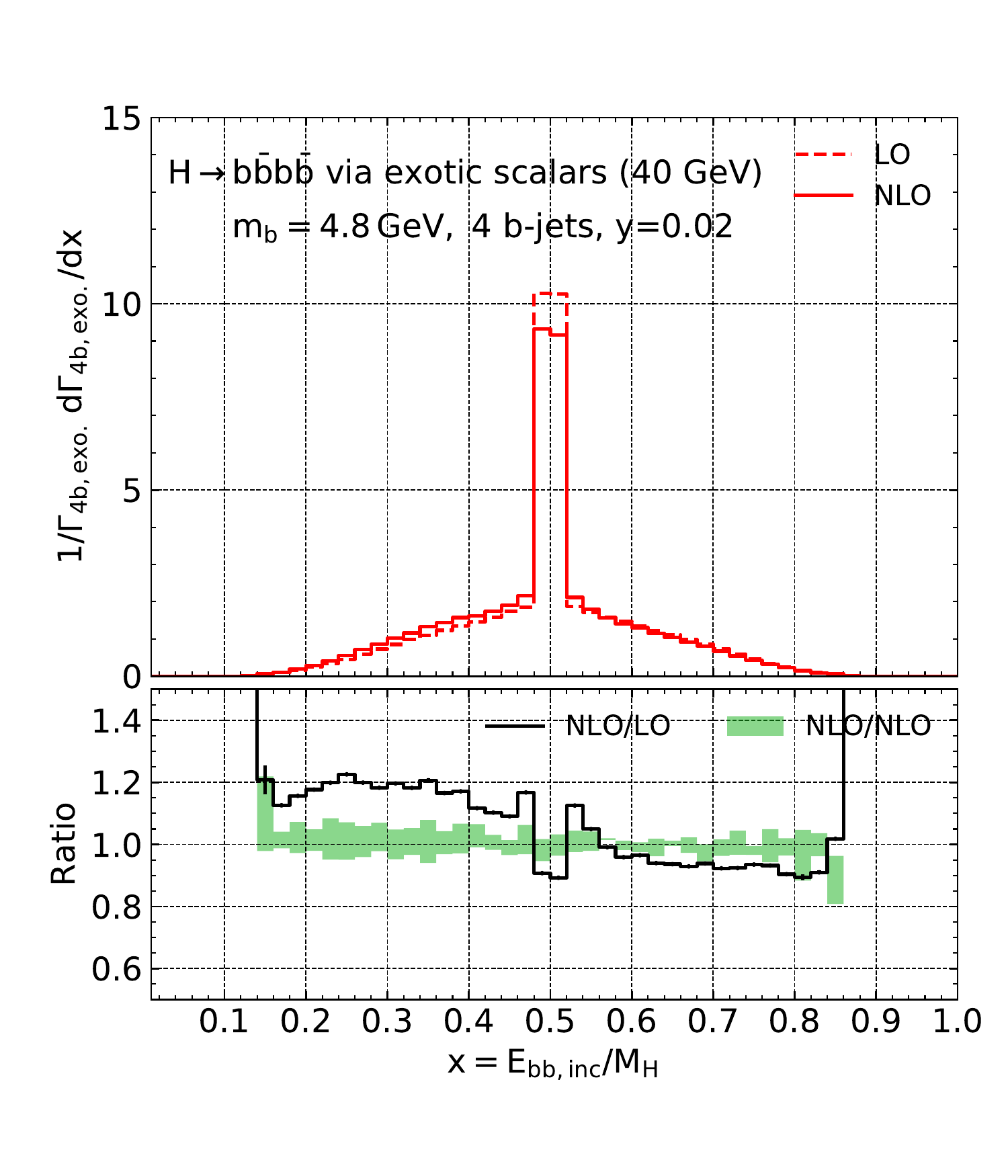}
\caption{
Similar to Fig.~\ref{fig:Ambbi} for distribution as a function of
the inclusive energy of all $b$-jet pairs.
\label{fig:Aebbi}}
\end{figure}

\begin{figure}[ht]
\centering
\includegraphics[width=0.47\textwidth]{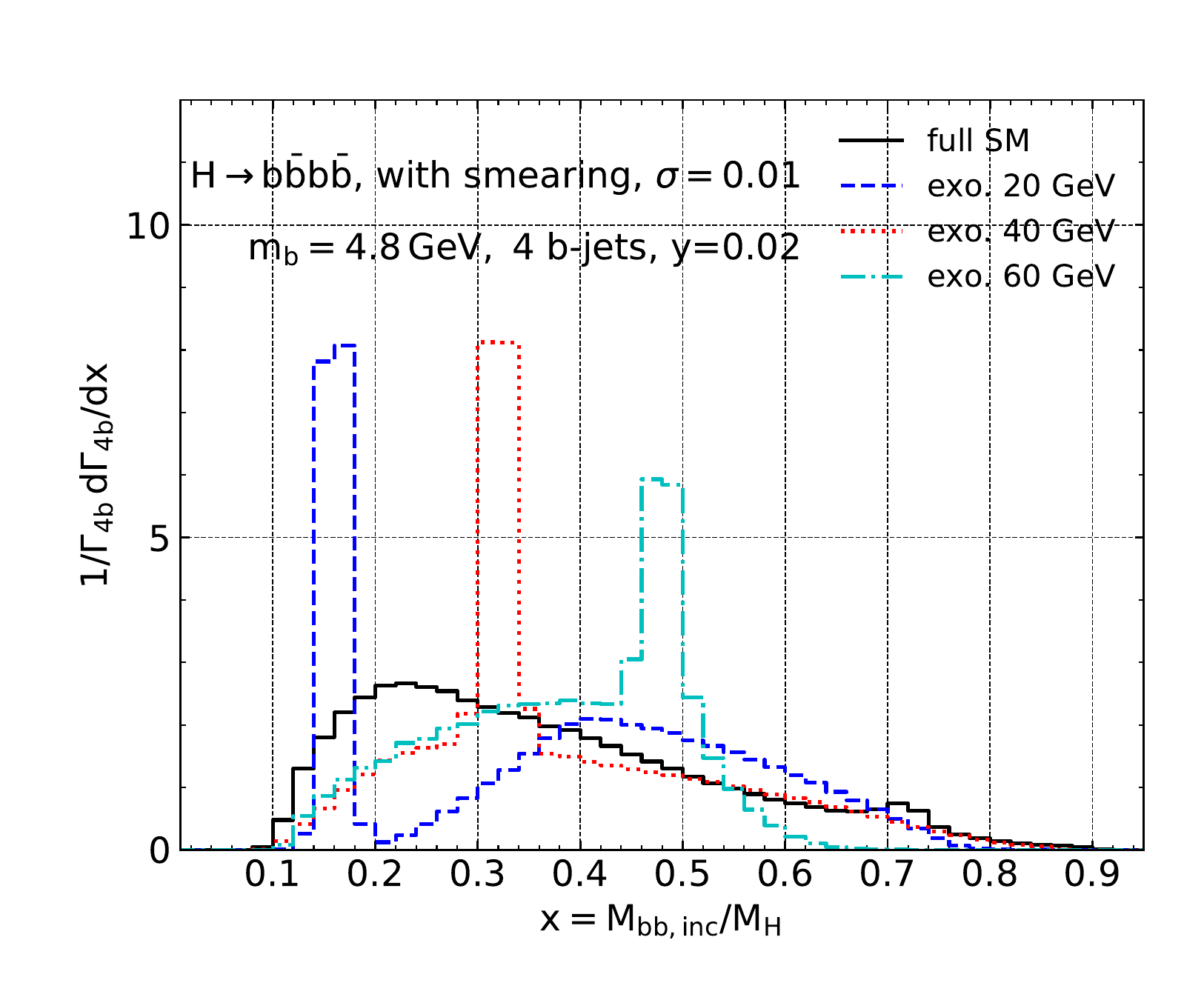}\hspace{0.1in}
\includegraphics[width=0.47\textwidth]{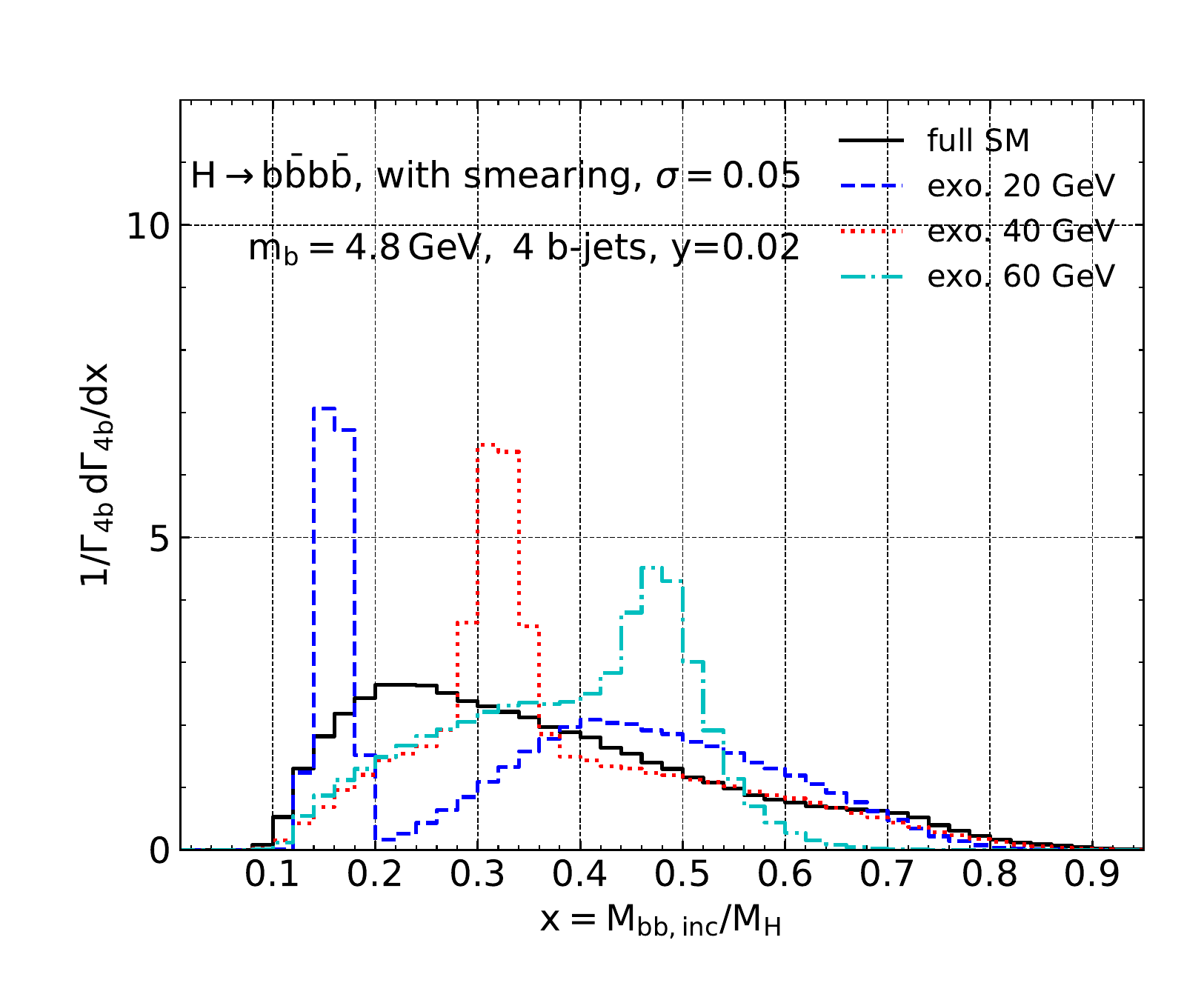}
\caption{
Comparison on normalized distribution of the Higgs boson decay to four bottom quarks as a function
of the inclusive invariant mass of all $b$-jet pairs, for the SM decay and exotic decay
via new light scalars with different masses, assuming an energy resolution
of 1\% (left plot) and 5\% (right plot), at NLO in QCD. 
\label{fig:Fmbbi}}
\end{figure}

\begin{figure}[ht]
\centering
\includegraphics[width=0.47\textwidth]{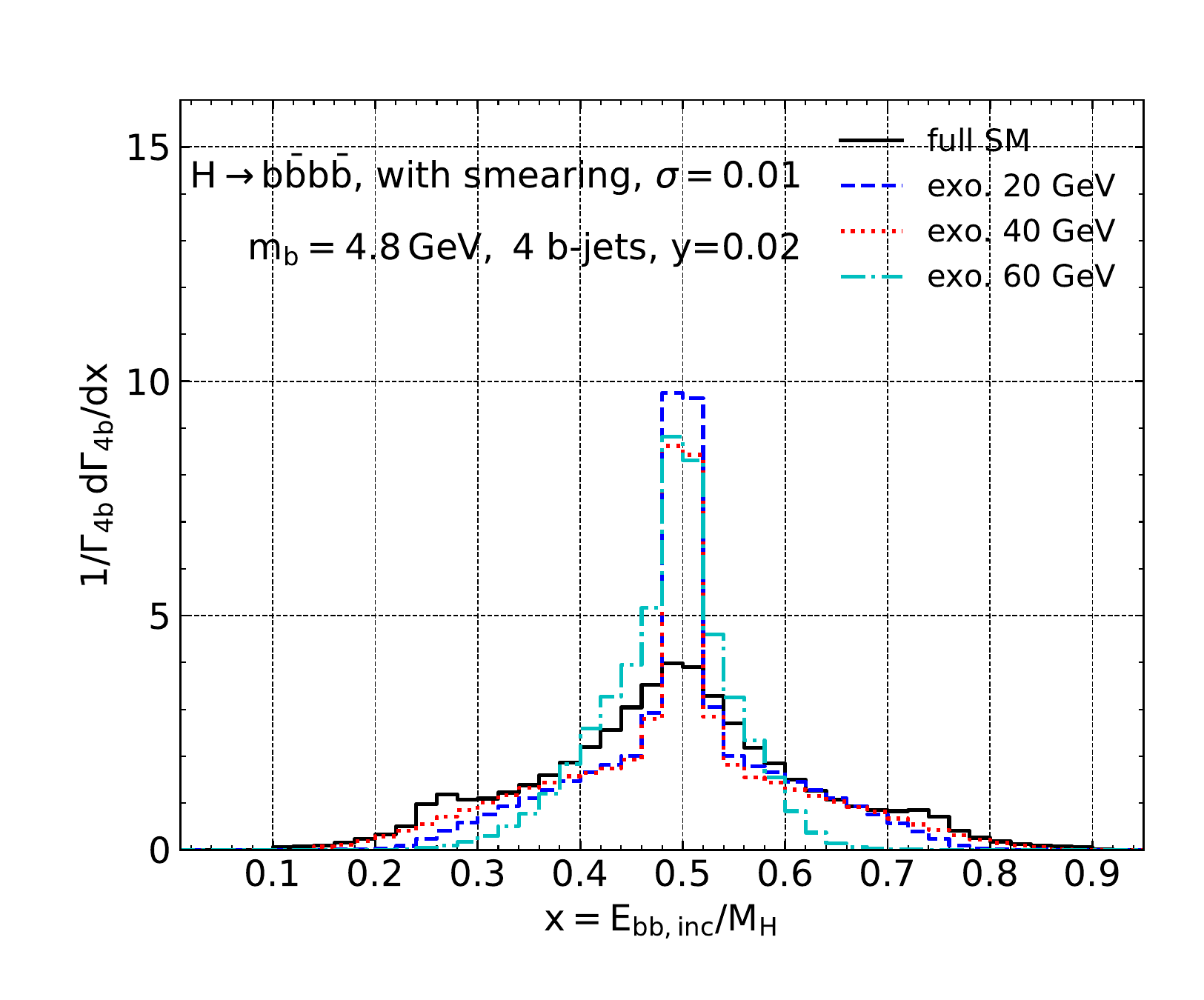}\hspace{0.1in}
\includegraphics[width=0.47\textwidth]{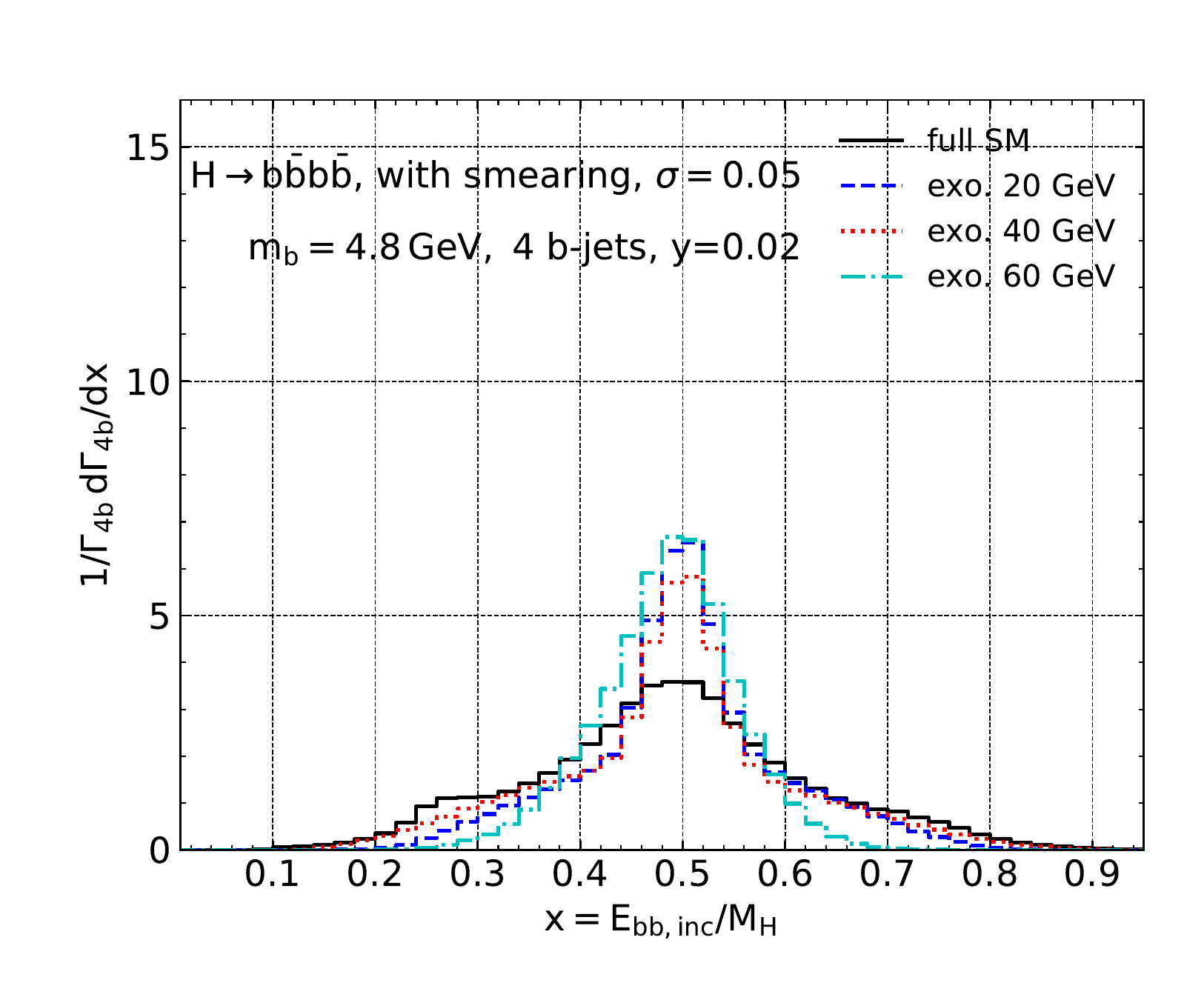}
\caption{
Similar to Fig.~\ref{fig:Fmbbi} for distribution as a function of
the inclusive energy of all $b$-jet pairs.
\label{fig:Febbi}}
\end{figure}

\section{Exotic decay}\label{sec:exo}

There have been proposals to study various exotic decay channels of the
Higgs boson at both the LHC~\cite{1312.4992} and future electron-positron colliders~\cite{Liu:2016zki}.
Higgs boson decays into four bottom quarks is among them and has
been employed to search for possible new scalars with mass $\lesssim m_H/2$
at the LHC~\cite{1806.07355}, as demonstrated by Feynman diagram in Fig.~\ref{fig:ff3}.
The ATLAS collaboration recently reported an upper limit of about 50\% on
the decay branching ratio of the Higgs boson to four bottom quarks via
two light scalars~\cite{1806.07355}, depending on the mass and lifetime of the new scalar. 
In this section we first investigate the QCD corrections to this exotic
decay channel as we did for the SM case.
Afterwards we compare various kinematic distributions for Higgs boson decays into
four bottom quarks via two light scalars to those predicted in the SM.  

\subsection{Kinematic distributions and QCD corrections}

In the calculation of the exotic decay, we assume the light scalar to be CP-even
similar to the SM Higgs boson.
We renormalize the Yukawa coupling between the bottom quark and the new scalar
with the $\overline{\rm MS}$ scheme. 
We also assume the new scalar has a small width thus the narrow width approximation
can be applied.
We are mostly interested in two kinematic distributions, i.e., the inclusive invariant
mass distribution and the inclusive energy distribution of $b$-jet pairs.
They are relevant for the exotic decay via new scalars which will show up
either as a sharp peak at the scalar mass or a sharp peak at half of the Higgs mass
respectively.
We plot the normalized distribution of $M_{b\bar b, inc}$ in Fig.~\ref{fig:Ambbi} for
a scalar mass of 20 and 40 GeV.
We compare the shape at LO and NLO.
For a lower mass of the new scalar, the QCD corrections barely modify the shape
except for the tail regions.
We observe a slightly broader peak with the QCD corrections for a larger scalar mass.
The impact of QCD corrections are similar for $E_{b\bar b, inc}$ shown in
Fig.~\ref{fig:Aebbi}.

\subsection{Comparison to the SM case}

For a future electron-positron collider, for instance, CEPC, the Higgs
boson can be produced abundantly and collected with a high efficiency.
With a center of mass energy of 250 GeV and an integrated luminosity of 5.6 ab$^{-1}$,
we expect a total number of the Higgs boson of about $1.1\times 10^6$ in $ZH$ production.
Thus in the SM it predicts a total of about 4000 events for the Higgs boson decaying into
four bottom quarks based on the decay branching ratio in Sect.~\ref{sec:sm}.
Such a rate can be measured with a precision of 2\% if taking into account only
statistical uncertainties and assuming perfect background rejection.
The SM decay of course contributes as an important background for searches of
the exotic decay via two new scalars, not mentioning other non-resonant processes,
e.g., $e^+e^-\rightarrow Zb\bar b b\bar b$.
In Fig.~\ref{fig:Fmbbi} we compare the normalized distribution of $M_{b\bar b, inc}$
for the decay to four bottom quarks in the SM and via the new scalars at 
NLO in QCD.
The SM result includes both decays via Yukawa couplings and electroweak
couplings.
We consider new scalars with a mass of 20, 40 and 60 GeV respectively.
We further apply a Gaussian smearing on the distributions since the signal over
background ratio strongly depends on the detector energy resolution.
In the left and right plots we assume an energy resolution of 1\% and 5\%
respectively.
One can see clearly distortions of the signal peak while the background is
less modified.
In Fig.~\ref{fig:Febbi} we present similar comparison for normalized
distribution of $E_{b\bar b, inc}$.
Impact of the energy smearing are similar as shown in Fig.~\ref{fig:Fmbbi}.
The inclusive energy of $b$-jet pairs show less discrimination power as comparing to
the inclusive invariant mass of $b$-jet pairs especially after taking into account the realistic
jet energy resolution.  	  

\section{Conclusions}\label{sec:con}

We study in details decays of the standard model Higgs boson to two bottom quark
and anti-quark pairs.
The hadronic decay can be triggered either by Yukawa couplings of
bottom and top quarks or the electroweak couplings.
Both channels are calculated to next-to-leading order in QCD with the former one
utilizing effective theory with top quarks integrated out.
We found large NLO QCD corrections for decay via Yukawa couplings that lead
to a 90\% increase of the inclusive partial decay width and residual QCD scale
variations of about 20\%.
On another hand we found moderate NLO QCD corrections of about 12\% for 
inclusive partial decay width of decay via electroweak couplings.
We also calculated various jet cross sections and kinematic distributions.
The QCD corrections can result in significant changes not only on
normalizations but also on shape of various distributions.
At future Higgs factory, all hadronic decay channels of the Higgs boson
including the one to four bottom quarks can be explored and measured.
Especially they can be employed to directly search for new physics
beyond the standard model, e.g., possible new light scalars that can
induce exotic decay of the Higgs boson to four bottom quarks.
Thus we further compare predictions on various kinematic distributions
of the decay in the SM and induced by those new scalars.
To complete the study on searches for new physics in the four
bottom quark decay channel it will be desirable to carry out a refined
study with all SM backgrounds included, e.g., those from non-Higgs boson production
processes, and with realistic detector coverage and resolution included.
Further matching of the NLO calculations with parton showering and QCD
hadronizations will also be required.
We leave those topics for a future study.

\section*{Acknowledgments}

The work of J.~Gao was sponsored by CEPC theory program and by the National Natural
Science Foundation of China under the Grant No. 11875189 and No.11835005.
The author would like to thank Qing Hong Cao, Jian Wang, Li Lin Yang, Hao Zhang and Hua Xing Zhu
for useful discussions, and thank G. Heinrich for clarifications on implementation of
Higgs effective coupling in GoSam 2.0.

\end{document}